\documentclass[11pt,a4paper]{article}
\usepackage[T1]{fontenc}
\usepackage[utf8]{inputenc}
\usepackage[english]{babel}
\usepackage[a4paper,margin=2.20cm]{geometry}
\usepackage{lmodern,microtype}
\usepackage{amsmath,amssymb,mathtools,bm}
\usepackage{booktabs,tabularx,array,longtable}
\usepackage{caption}
\usepackage{graphicx,float}
\usepackage[section]{placeins}
\usepackage{authblk}
\usepackage{xcolor}
\usepackage{enumitem}
\usepackage{cite}
\usepackage{xurl}
\usepackage[unicode]{hyperref}
\hypersetup{colorlinks=true,linkcolor=blue!45!black,citecolor=blue!45!black,urlcolor=blue!55!black,
 pdftitle={Rank and Scale Tests of Shell-Motivated Fermion-Mass Ansatze with a Constrained Neutrino Extension},
 pdfauthor={Yaroslav D. Krivenko-Emetov and Oleksii M. Lytvynenko}}
\newcommand{\MSbar}{\overline{\mathrm{MS}}}
\newcommand{\Rone}{R_1}
\newcommand{\Rtwo}{R_2}
\newcommand{\Rthree}{R_{\nu\ell}}
\newcommand{\Rhyper}{R_H}
\newcommand{\RBu}{R_B}
\newcommand{\RuL}{R_{uL}}

\newcommand{\sth}{s_{\mathrm{th}}}
\newcommand{\GeV}{\mathrm{GeV}}

\newcommand{\eV}{\mathrm{eV}}

\newcommand{\diag}{\operatorname{diag}}
\newcommand{\trans}{\mathsf{T}}
\newcommand{\dd}{\mathrm{d}}
\newcommand{\dperp}{d_{\perp}}

\title{\textbf{Rank and Scale Tests of Shell-Motivated Fermion-Mass Ansatze with a Constrained Neutrino Extension}}
\author{%
\textbf{Yaroslav D. Krivenko-Emetov\textsuperscript{1,2,3,*}}\quad and \quad\textbf{Oleksii M. Lytvynenko\textsuperscript{2,a}}\par\medskip
{\small\itshape \textsuperscript{1}National Technical University of Ukraine ``Igor Sikorsky Kyiv Polytechnic Institute'', Kyiv, Ukraine}\\
{\small\itshape \textsuperscript{2}Institute for Nuclear Research, National Academy of Sciences of Ukraine, Kyiv, Ukraine}\\
{\small\itshape \textsuperscript{3}Taras Shevchenko National University of Kyiv, Kyiv, Ukraine}%
}
\date{}

\begin{document}
\sloppy
\maketitle
\begingroup
\renewcommand{\thefootnote}{\fnsymbol{footnote}}
\footnotetext[1]{Corresponding author: \href{mailto:y.krivenko-emetov@kpi.ua}{y.krivenko-emetov@kpi.ua}; \href{mailto:krivemet@ukr.net}{krivemet@ukr.net}.}
\endgroup
\begingroup
\renewcommand{\thefootnote}{\alph{footnote}}
\footnotetext[1]{Former employee of the Institute for Nuclear Research, National Academy of Sciences of Ukraine.}
\endgroup

\begin{abstract}
We analyze logarithmic ans\"atze for the nine charged-fermion Yukawa eigenvalues, all evaluated in the same renormalization scheme at the common scale $\mu=M_Z$, where $M_Z$ is the mass of the $Z$ boson.  Matrix rank and left-null spaces are used to identify parameter-independent relations before any numerical fit is performed.  An exhaustive scan of 52 distinct feature planes in three coefficient blocks gives 156 rank-eight charged-sector models.  The smallest residual at fixed $M_Z$ belongs to the slope model $M_B$, with $\chi^2_{\min}=0.00119$ when an illustrative 25\% model discrepancy is assigned to each state.  This model has only one parameter-independent relation: it connects the first-to-third-generation slopes, while all three middle-generation masses are still interpolated rather than predicted.

We therefore audit seven- and six-parameter descendants of $M_B$.  The leading seven-parameter curvature-plane model, $M_{B+dLR}$, has two charged-sector constraints, $\chi^2_{\min}=0.01336$, and a maximum conditional shift of $1.99\%$, but it leaves the strange-quark mass $m_s$ unidentified in leave-one-out (LOO) reconstruction.  Here LOO means that one mass is removed, the remaining eight masses are fitted, and the omitted mass is reconstructed.  The simplest complete six-parameter model is $M_{B+\ell T}$, with sector curvature $A_s\propto P_u-3Y_L=2P_\ell+T_{3L}$.  The projectors $P_u$ and $P_\ell$ select the up-quark and charged-lepton sectors, while $Y_L$ and $T_{3L}$ denote left-handed hypercharge and weak isospin.  This model has 9/9 LOO coverage, a maximum conditional shift of $6.81\%$, and a worst LOO error of $13.38\%$.  The fixed integer family $A_\ell:A_d:A_u=(3n+2):-(n+1):n$ retains the same parameter count.  The choices $n=2$ and $n=3$ reduce the worst LOO error to $7.71\%$ and $4.99\%$, respectively.  Nested selection over $1\le n\le15$ chooses $n=2$ in eight of nine folds, making it the conservative predictive choice; $n=3$ is the central-data minimax choice.

The targeted six-parameter rule $M_{B+dLR}^{s14}$, $A_d=B_d/14$, reconstructs $m_s$ with a $0.51\%$ LOO error but has a $25.06\%$ worst-case error over all nine states.  A focused follow-up audit of the otherwise poor blockwise descendant $M_{2+B+dLR}$ identifies three much better fixed-scale repairs: $I_\ell-I_d=A_\ell/3$, a shifted-center rule $I_\ell-I_d+(B_\ell-B_d)/3=0$, and the quantum-defect form $I_\ell-I_d+B_d/16=0$.  They retain six parameters and 9/9 LOO coverage, with maximum conditional shifts $1.86\%$, $1.91\%$, and $1.95\%$, and worst LOO errors $3.82\%$, $5.78\%$, and $3.91\%$, respectively.  Under the stated Dirac-like bookkeeping, all four $dLR$-based six-parameter continuations give vanishing neutrino curvature, $A_\nu=0$, and therefore a neutrino-mass sum $\sum m_\nu=0.060504\,\eV$ only for normal ordering (NO).  The $M_{B+\ell T_n}$ family instead gives $A_\nu=A_\ell$ and therefore $\sum m_\nu=0.058914\,\eV$ for NO or $0.103329\,\eV$ for inverted ordering (IO), without an additional continuous parameter.

We then construct a minimal operator-matching completion of the integer family.  Two gauge-invariant logarithmic matching insertions, ${\cal O}_0$ and ${\cal O}_1$, have charged-sector eigenvalues
\(q_0=(3,-1,1)\) and \(q_1=(2,-1,0)\).  Their signs follow from the mass-independent quantum-number identities
\(Q_0=2(P_u-3Y_L)\) and \(Q_1=-3(B-L)-P_\ell+P_u\), rather than from the fermion masses.  If the leading insertion has two or three symmetry-related contractions while the correction has one, the matching coefficient is fixed before the spectrum is used to $C_1/C_0=1/2$ or $1/3$.  These are explicit conditional ultraviolet (UV) classes, not yet consequences of the published $SU(15)_p$ dynamics, whose nonperturbative reduced coefficients remain independent.  A two-loop Standard Model (SM) renormalization-group (RG) running audit also changes the interpretation of the targeted rule: $A_d/B_d$ is close to $1/15$ near $6.1\times10^6\,\GeV$ and evolves to $0.070795$ at $M_Z$, close to $1/14$.  Thus $1/14$ may be an infrared mnemonic for an RG-evolved $1/15$ matching factor, provided a microscopic theory independently predicts the matching scale and threshold corrections.  The resulting operator and RG construction is more restrictive than a numerical coincidence, but it remains a falsifiable matching hypothesis rather than experimental confirmation of preons.
\end{abstract}


\subsection*{Notation and reading guide}
The notation used repeatedly below is collected here so that each later model can be read without reconstructing definitions from several sections.  The sector label is $s\in\{\ell,d,u,\nu\}$ for charged leptons, down-type quarks, up-type quarks, and neutrinos; the generation label is $g\in\{1,2,3\}$.  All charged Yukawa eigenvalues $y_{s,g}$ are expressed in the $\MSbar$ scheme at the common scale $\mu=M_Z$, and $z_{s,g}\equiv\ln y_{s,g}$.  Each sector is decomposed as
\begin{equation}
z_{s,g}=I_s+B_s(g-2)+A_sh_g,\qquad h_g=(+1,-1,+1),
\label{eq:readertriplet}
\end{equation}
where $I_s$ is the logarithmic level, $B_s$ is the first-to-third-generation slope, and $A_s$ is the discrete curvature that locates the middle generation relative to the endpoints.  Thus a restriction on $B_s$ does not by itself predict a middle-generation mass, whereas a restriction on $A_s$ does.

The principal diagnostics also have distinct meanings.  The quantity $\chi^2_{\min}$ is a conditional goodness-of-fit statistic computed with the explicitly stated model discrepancy $\sth$; it is not a discovery significance.  The maximum conditional shift is the largest fractional adjustment needed to place the nine central values on the model subspace.  LOO denotes leave-one-out reconstruction: one mass is omitted, the model is fitted to the other eight, and the omitted mass is predicted.  ``9/9 LOO coverage'' means that this reconstruction is defined for every charged state.  NO and IO denote normal and inverted neutrino mass ordering.  RG, UV, CKM, PMNS, and FRG denote renormalization group, ultraviolet, Cabibbo--Kobayashi--Maskawa, Pontecorvo--Maki--Nakagawa--Sakata, and functional renormalization group, respectively.  Table~\ref{tab:modelmap} later gives the complete definitions and predictive status of the named models.

\section{Introduction}\label{sec:intro}
The Standard Model (SM) describes known gauge interactions with exceptional precision, but its Yukawa matrices and mixing parameters remain empirical inputs.  The origin of quark and lepton masses, their strong hierarchies, and their contrasting mixing patterns is therefore a central part of the flavor problem \cite{Feruglio2015,Xing2020,DingKing2024,DingValle2025}.  Representative mechanisms include Abelian selection rules \cite{FroggattNielsen1979}, partial compositeness \cite{Kaplan1991}, modular and discrete symmetries \cite{DingKing2024,DingValle2025}, radiative rank generation \cite{MohantaPatel2022,Jana2025,MohantaPatel2026}, and multi-Higgs or family-gauge constructions \cite{Baek2024}.  Empirical mass relations, most famously the Koide relation, provide complementary diagnostics but do not by themselves identify a microscopic mechanism \cite{Koide1983,XingZhang2006,GaoLi2016}.

Preon or composite-fermion models suggest a more radical possibility: the observed generations may be distinct low-energy states of confined constituents \cite{Harari1979,Shupe1979}.  Such a proposal must do far more than fit nine numbers.  It must produce light chiral fermions, satisfy anomaly matching, suppress unwanted low-energy operators, account for mixing, and respect collider limits on compositeness \cite{tHooft1980,AssiEtAl2026,PDG2026,CMSTauStar2025}.  Recent work addresses several of these structural obstacles.  A chiral $SU(15)_p$ theory produces three generations of composite quarks and leptons, a composite Higgs, and a benchmark that fits all charged masses and CKM mixing with flavor-breaking spurions \cite{Dobrescu2022,AssiDobrescu2025,AssiEtAl2026}.  A functional renormalization-group (FRG) study of Bars--Yankielowicz chiral gauge theories finds a regime of confinement without dynamical symmetry breaking, supporting the possibility of light anomaly-matching composite fermions \cite{LiPastorVatani2026}.  A complementary $SL(16,\mathbb C)$ construction uses anomaly matching to select an $SU(8)$ hyperflavor sector containing three chiral quark--lepton families \cite{Chkareuli2024}.  These results demonstrate that relevant ingredients can coexist within specific constructions; they are not experimental evidence for substructure.  The present work asks a narrower infrared question: what exact, parameter-independent statements follow from a specified shell-motivated mass ansatz when all charged inputs are defined in one scheme at one scale?

A finite self-consistent system with a regular discrete spectrum can possess shell corrections, as in nuclear macroscopic--microscopic theory and semiclassical spectral analysis \cite{Strutinsky1967,Strutinsky1968,Gutzwiller1971}.  Here this analogy motivates a first discrete harmonic, not a calculated Strutinsky correction.  With only three observed generations, $(+1,-1,+1)$ is not itself orthogonal to the constant vector, but, after its constant component is absorbed into the intercept, it spans the unique direction complementary to a constant and a linear trend.  The corresponding data contrast is proportional to $(+1,-2,+1)$.  The term ``shell-motivated'' therefore describes a possible interpretation of the basis, not established constituent dynamics.

The smooth log-linear trend is separately motivated by a conditional dimensional-transmutation argument.  The confining scale of an asymptotically free theory depends exponentially on the inverse ultraviolet coupling.  A weak systematic change of an effective group factor between bound-state channels can therefore generate an approximately exponential hierarchy.  This derivation is retained because it explains why a linear function of the generation label may be natural in logarithmic mass space, but every model-dependent premise is stated explicitly.

The original M1 and M2 constructions, together with $M_B$, $M_{B+uL}$, $M_H$, $M_{dG}$, $M_{B,LR}$, and the unsuccessful auxiliary descendants, are retained as the historical genealogy and reproducibility controls of the scan.  Their definitions and diagnostic numbers are preserved, but they are no longer treated as coequal physical candidates.  The substantive comparison instead focuses on the complete six-parameter $M_{B+\ell T_n}$ family, the targeted $M_{B+dLR}^{s14}$ rule, and the best repaired $dLR$ descendants.  This reorganization separates five questions: how accurately the endpoint slopes are related; whether the middle generation enters a null-space constraint; whether an omitted mass can actually be reconstructed; whether a neutral continuation is kinematically viable; and whether the surviving integer coefficients can be obtained from quantum numbers and a pre-spectral microscopic matching rule.

The manuscript makes the following distinctions throughout:
\begin{enumerate}[leftmargin=*,itemsep=2pt]
\item exact rank statements versus conditional goodness-of-fit numbers;
\item a phenomenological basis versus a microscopic shell calculation;
\item parameter-elimination invariants at fixed $\mu$ versus true RG invariants;
\item oscillation mass splittings versus three independently measured absolute neutrino masses;
\item exploratory post-selection versus out-of-sample prediction.
\end{enumerate}

The presentation follows that logic.  Section~\ref{sec:model} first defines the three spectral coefficients and records the historical model genealogy before introducing the inputs and fit convention.  Section~\ref{sec:chargedresults} ranks the charged relations, and Sec.~\ref{sec:neutralresults} separately states the neutral matching assumptions.  Section~\ref{sec:jointaudit} combines both criteria, identifies the predictive integer rays and targeted-$s$ alternative, constructs the two microscopic operator images, and performs their matching and RG audit.  Only then do Secs.~\ref{sec:prior} and \ref{sec:limits} compare the conditional completion with current preon dynamics and state what remains unproved.

\section{Model and methods}\label{sec:model}
\subsection{A shell-motivated discrete-curvature basis}
For a finite system one may schematically decompose an energy as
\begin{equation}
E=E_{\rm smooth}+\delta E_{\rm shell},
\end{equation}
where the separation depends on a specified smoothing prescription.  Periodic orbits and repeated level bunching can generate oscillatory contributions to the spectral density.  Retaining only a first harmonic motivates
\begin{equation}
h_g=\cos[\pi(g-1)]=( +1,-1,+1),\qquad g=1,2,3.
\label{eq:hg}
\end{equation}
No microscopic spectrum, smoothing width, plateau condition, or constituent Hamiltonian is calculated here.  Equation~\eqref{eq:hg} is used as a phenomenological basis function.  For an isolated triplet, a constant, a linear slope, and $h_g$ form a saturated three-dimensional basis.  Predictive content can arise only by sharing selected coefficients across sectors.

\subsection{How to read the three coefficients}\label{sec:readcoeff}
The generic triplet form used below is $z_{s,g}=I_s+B_s(g-2)+A_sh_g$, where $z_{s,g}=\ln y_{s,g}$.  Its meaning can be made completely explicit.  For any three measured values in one sector,
\begin{equation}
 I_s=\frac{z_{s,1}+2z_{s,2}+z_{s,3}}{4},\qquad
 B_s=\frac{z_{s,3}-z_{s,1}}{2},\qquad
 A_s=\frac{z_{s,1}-2z_{s,2}+z_{s,3}}{4}.
\label{eq:inverseIBA}
\end{equation}
Thus $I_s$ fixes the logarithmic level, $B_s$ is the endpoint slope, and $A_s$ is one quarter of the discrete second difference.  Equivalently,
\begin{equation}
 Q_s\equiv\frac{y_{s,1}y_{s,3}}{y_{s,2}^2}=e^{4A_s}.
\label{eq:Qcurvature}
\end{equation}
The curvature has an immediate interpretation.  If the three masses followed an exact geometric progression, then $Q_s=1$ and $A_s=0$.  Values $Q_s<1$ and $Q_s>1$ describe opposite departures of the middle generation from that smooth progression.  At $M_Z$ the direct charged values are
\begin{equation}
 (A_\ell,A_d,A_u)=(-0.62946,\ 0.24653,\ -0.15537),\qquad
 (Q_\ell,Q_d,Q_u)=(0.080633,\ 2.68081,\ 0.537154).
\label{eq:directcurvatures}
\end{equation}
This also clarifies the limitation of the shell language.  For exactly three points, the vector $(+1,-1,+1)$ does not reveal an independently resolved oscillation: after removing a constant and a slope, it is simply the unique remaining curvature direction.  A microscopic shell claim would require additional levels or an independently specified constituent Hamiltonian.

\subsection{What a named model means in this analysis}\label{sec:modelmeaning}
A ``model'' below is first of all a precisely defined linear subspace of the nine charged log-eigenvalues, not yet a complete field theory.  The generic triplet form written immediately above has the nine independent charged coefficients
\begin{equation}
 (I_\ell,I_d,I_u;\ B_\ell,B_d,B_u;\ A_\ell,A_d,A_u)
\end{equation}
and therefore interpolates all nine charged masses.  A predictive charged model is obtained only by sharing coefficients between sectors or by restricting one coefficient block to a lower-dimensional feature plane.  Its physical statement is then read block by block:
\begin{itemize}[leftmargin=*,itemsep=2pt]
\item a restriction on $I_s$ concerns the overall logarithmic level of a sector;
\item a restriction on $B_s$ concerns the first-to-third-generation hierarchy and does not by itself test the middle mass;
\item a restriction on $A_s$ concerns the discrete curvature and therefore directly tests the position of the middle generation relative to the two endpoints.
\end{itemize}
The rank and left-null space specify what the model actually predicts.  A small residual then says that the measured spectrum lies close to that subspace; it does not by itself identify the microscopic dynamics that generated it.  The named models are introduced in the order M1 $\rightarrow$ M2 $\rightarrow$ rank-eight projector restrictions $\rightarrow$ the post-selected rank-seven synthesis.  Neutral extensions are separate matching hypotheses and are never inferred from a good charged fit alone.

\subsection{Conditional RG motivation of the exponential trend}\label{sec:rgmot}
Let a hypothetical confining gauge group $G_P$ be asymptotically free.  For $\alpha_P=g_P^2/(4\pi)$, the one-loop equation is
\begin{equation}
\mu\frac{\dd\alpha_P}{\dd\mu}=-\frac{b_P}{2\pi}\alpha_P^2+\mathcal O(\alpha_P^3),\qquad b_P>0,
\label{eq:beta}
\end{equation}
with solution
\begin{equation}
\frac{1}{\alpha_P(\mu)}=\frac{1}{\alpha_P(M_*)}+\frac{b_P}{2\pi}\ln\frac{\mu}{M_*}.
\end{equation}
Dimensional transmutation then generates a strong scale \cite{GrossWilczek1973,Politzer1973}
\begin{equation}
\Lambda_P=M_*\exp\left[-\frac{2\pi}{b_P\alpha_P(M_*)}\right].
\label{eq:lambdaP}
\end{equation}
Equation~\eqref{eq:lambdaP} alone does not distinguish generations.  An additional model hypothesis is required.  Suppose the labels $g=1,2,3$ identify configurations $R_g$ with effective channel factors $c_g$ at $M_*$.  A simple schematic matching rule is
\begin{equation}
\alpha^{\rm eff}_{P,g}(M_*)=c_g\alpha_P(M_*),\qquad c_g=\frac{C_2(R_g)}{C_2(R_1)},
\label{eq:matching}
\end{equation}
where the second equality is an example, not a consequence of the data.  The channel-dependent scale is then
\begin{equation}
\Lambda_g=M_*\exp\left[-\frac{\mathcal A}{c_g}\right],\qquad \mathcal A\equiv\frac{2\pi}{b_P\alpha_P(M_*)}.
\end{equation}
If the effective factor varies weakly and systematically,
\begin{equation}
c_g=1+\epsilon(g-1),\qquad |\epsilon(g-1)|\ll1,
\end{equation}
then
\begin{align}
\ln\frac{\Lambda_g}{\Lambda_1}
&=\frac{\mathcal A\epsilon(g-1)}{1+\epsilon(g-1)}\\
&=\mathcal A\epsilon(g-1)-\mathcal A\epsilon^2(g-1)^2+\mathcal O(\epsilon^3).
\end{align}
To first order,
\begin{equation}
\boxed{\Lambda_g\simeq\Lambda_1\exp[\gamma(g-1)]},\qquad
\gamma=\frac{2\pi\epsilon}{b_P\alpha_P(M_*)}.
\label{eq:exptrend}
\end{equation}
If the smooth mass in sector $s$ scales as a fixed power of $\Lambda_{s,g}$, constants can be absorbed into a sector offset and slope,
\begin{equation}
\ln m^{\rm smooth}_{s,g}=I_s+B_s(g-2).
\label{eq:smooth}
\end{equation}
A multiplicative first-harmonic modulation,
\begin{equation}
m_{s,g}=m^{\rm smooth}_{s,g}\exp(A_sh_g),
\end{equation}
gives
\begin{equation}
\ln m_{s,g}=I_s+B_s(g-2)+A_sh_g.
\label{eq:saturated}
\end{equation}
Thus an approximately exponential smooth hierarchy can arise from dimensional transmutation if generations correspond to channels with systematically varying effective group characteristics.  This is a conditional motivation, not a microscopic derivation: $G_P$, the admissible $R_g$, $b_P$, $\epsilon$, the matching rule in Eq.~\eqref{eq:matching}, and the proportionality of mass to the confinement scale are not derived.  The mechanism must also be distinguished from the ordinary SM RG evolution of already matched Yukawa couplings, studied below as a scale-sensitivity test.

\subsection{Historical starting point: M1}
Let $C=0$ for leptons and $C=1$ for quarks, and define $T=2T_3=-1$ for charged leptons and down-type quarks, and $T=+1$ for neutrinos and up-type quarks.  The down-sector projector is
\begin{equation}
D(C,T)=C\frac{1-T}{2}.
\end{equation}
The initial exploratory ansatz was
\begin{align}
\ln y_{\rm M1}(g,C,T;\mu)=&\;\alpha_0+\alpha_C C+\alpha_TT\\
&+(\gamma_0+\gamma_CC+\gamma_TT)(g-1)\\
&+[A_0+A_dD(C,T)]h_g.
\label{eq:M1}
\end{align}
It contains eight coefficients.  Equivalent parameter-elimination ratios may be written in terms of running masses because their numerator and denominator have the same total mass dimension.

In the charged sector this notation has a simpler equivalent meaning.  The three functions $(1,C,T)$ resolve the three offsets $I_s$ and the three slopes $B_s$ independently, while $D=0$ for charged leptons and up quarks and $D=1$ for down quarks.  Consequently M1 imposes only
\begin{equation}
A_\ell=A_u,
\end{equation}
with $A_d$ independent.  Its physical hypothesis is therefore a common discrete-curvature contribution for the charged-lepton and up-quark triplets, not a universal prediction of all nine masses.  One might associate such equality with a shared constituent channel or shell correction, but no operator or symmetry enforcing it is supplied.

Order the charged states as
\begin{equation}
(e,d,u,\mu,s,c,\tau,b,t).
\label{eq:ordering}
\end{equation}
Its rank-eight design matrix gives the single relation
\begin{equation}
\boxed{\Rone(\mu)\equiv\frac{y_u(\mu)y_\mu^2(\mu)y_t(\mu)}{y_e(\mu)y_c^2(\mu)y_\tau(\mu)}=1.}
\label{eq:R1}
\end{equation}
The data give $\Rone(M_Z)=6.662$.  Its failure and alternating residuals motivated M2; M1 is not retained as a competitive model in the comparisons below.

\subsection{Historical residual-motivated baseline: M2}
M1 forces the charged-lepton and up-quark triplets to share one curvature.  The M1 residuals show opposite alternating patterns in those sectors.  Introduce mutually exclusive projectors
\begin{equation}
L=1-C,\qquad D=C\frac{1-T}{2},\qquad U=C\frac{1+T}{2},\qquad L+D+U=1.
\end{equation}
M2 is
\begin{align}
\ln y_{\rm M2}(g,C,T;\mu)=&\;\beta_0+\beta_TT
 +(\eta_0+\eta_CC+\eta_TT)(g-2)\\
&+(A_\ell L+A_dD+A_uU)h_g.
\label{eq:M2}
\end{align}
It again has eight parameters.  Its design matrix has rank eight and left-null vector
\begin{equation}
\bm w_2=(+1,-1,0,+2,-2,0,+1,-1,0)^{\trans},
\end{equation}
so the single relation is
\begin{equation}
\boxed{\Rtwo(\mu)\equiv\frac{y_e(\mu)y_\mu^2(\mu)y_\tau(\mu)}{y_d(\mu)y_s^2(\mu)y_b(\mu)}=1.}
\label{eq:R2}
\end{equation}
The up-type triplet is now interpolated exactly.  M2 is not an independently pre-registered alternative: its coefficient allocation was chosen after inspecting M1 residuals.

Again the charged-sector content is simpler than the original notation suggests.  The three slope coefficients span independent $B_\ell,B_d,B_u$, and the projectors $L,D,U$ make the three curvatures independent.  The only remaining restriction is
\begin{equation}
I_\ell=I_d
\end{equation}
at the fixed center $g=2$.  Thus M2 compares the centered logarithmic levels of the charged-lepton and down-quark triplets after their slopes and curvatures have been removed.  This may be read as an effective matching condition between two sectors, but its dependence on the chosen generation origin and its residual-driven construction prevent a symmetry interpretation at present.

\subsection{Systematic rank-eight projector family}\label{sec:projectorfamily}
The saturated charged-sector basis is
\begin{equation}
\ln y_s(g)=I_s+B_s(g-2)+A_sh_g,\qquad s\in\{\ell,d,u\}.
\label{eq:charged_sat}
\end{equation}
Its nine coefficients interpolate the nine charged Yukawa eigenvalues.  A rank-eight reduction is obtained by restricting exactly one block $C_s\in\{I_s,B_s,A_s\}$ to a two-dimensional feature plane,
\begin{equation}
C_s=c_1q_{1,s}+c_2q_{2,s},
\label{eq:featureplane}
\end{equation}
while the other two blocks remain sector resolved.  We scanned all independent pairs built from
\begin{equation}
1,\ P_\ell,\ P_d,\ P_u,\ Y_R,\ |Y_R|,\ Y_R^2,\ Y_L,\ |Y_L|,\ Y_L^2,
\ T_{3L},\ B-L,\ P_C,\ Y_L^2+Y_R^2,\ Y_LY_R.
\label{eq:featurelist}
\end{equation}
Here every entry in Eq.~\eqref{eq:featurelist} is a fixed, dimensionless sector feature, not an additional fitted parameter.  The symbol $1$ denotes the constant feature.  The indicator projectors $P_\ell$, $P_d$, and $P_u$ select, respectively, the charged-lepton, down-type-quark, and up-type-quark sectors; in the charged ordering $(\ell,d,u)$ their rows are
\begin{equation}
 P_\ell=(1,0,0),\qquad P_d=(0,1,0),\qquad P_u=(0,0,1).
\label{eq:sectorprojectors}
\end{equation}
The color-sector projector $P_C=P_d+P_u=(0,1,1)$ distinguishes quarks from leptons, so $P_\ell=1-P_C$ on these three charged sectors.  The quantities $Y_L$ and $Y_R$ are the Standard-Model hypercharges of the left- and right-chiral fermions in the convention $Q=T_{3L}+Y$; $T_{3L}$ is the third component of weak isospin of the left-handed state; and $B-L$ is baryon number minus lepton number.  Their charged-sector values are
\begin{equation}
\begin{array}{c|ccc}
 & \ell & d & u\\ \hline
Y_L & -\tfrac12 & \tfrac16 & \tfrac16\\
Y_R & -1 & -\tfrac13 & \tfrac23\\
T_{3L} & -\tfrac12 & -\tfrac12 & \tfrac12\\
B-L & -1 & \tfrac13 & \tfrac13
\end{array}.
\label{eq:chargedfeaturerows}
\end{equation}
Absolute values and squares, such as $|Y_R|$ and $Y_L^2$, remove the sign and test whether a coefficient follows a positive charge magnitude; $Y_L^2+Y_R^2$ is the corresponding quadratic chiral-charge strength.  By contrast, $Y_LY_R$ retains the relative sign and therefore tests a left--right-sensitive pattern.  In Eq.~\eqref{eq:featureplane}, $q_{a,s}$ means the value of the $a$th chosen feature in sector $s$, while $c_1$ and $c_2$ are the two fitted amplitudes spanning that feature plane.  Thus the quantum numbers are fixed inputs, whereas only the amplitudes are estimated from the masses.
After feature pairs spanning the same charged-sector plane are identified, the audit contains 52 planes and 156 rank-eight block models.  Every model has one charged left-null relation.  This exhaustive ranking remains exploratory because the same nine masses are used for selection and evaluation.

\subsection{Road map: definitions and physical content of the named models}\label{sec:modelmap}
Table~\ref{tab:modelmap} collects every named model before the numerical results are discussed.  The ``possible physical reading'' column is deliberately weaker than a derivation: it states what kind of dynamics could produce the restriction, while the last sentence in each entry states what has actually been established here.

\begingroup
\footnotesize
\setlength{\LTpre}{2pt}
\setlength{\LTpost}{4pt}
\begin{longtable}{@{}>{\raggedright\arraybackslash}p{1.25cm}>{\raggedright\arraybackslash}p{5.0cm}>{\raggedright\arraybackslash}p{3.4cm}>{\raggedright\arraybackslash}p{5.4cm}@{}}
\caption{Definitions, tested spectral content, and possible physical reading of the models.  $S_9$ denotes the saturated nine-parameter charged reference.}\label{tab:modelmap}\\
\toprule
model & mathematical definition & what is tested & possible physical reading and present status\\
\midrule
\endfirsthead
\toprule
model & mathematical definition & what is tested & possible physical reading and present status\\
\midrule
\endhead
$S_9$ & independent $(I_s,B_s,A_s)$ for $s=\ell,d,u$ & nothing; all nine charged inputs are interpolated & bookkeeping basis that separates level, endpoint slope, and discrete curvature\\
M1 & offsets and slopes affine in $(C,T)$; curvature $A_0+A_dD(C,T)$ & one charged relation $R_1=1$ & first attempt to encode color/isospin channels and one shell-like correction; rejected as a competitive fit\\
M2 & $I_\ell=I_d$; all three $B_s$ and $A_s$ are sector resolved & one charged relation $R_2=1$ & equality of the centered lepton and down-quark levels; chosen from the M1 residual pattern, not derived from symmetry\\
$M_B$ & $B_s\in\mathrm{span}(|Y_R|,Y_L)$; $I_s,A_s$ free & endpoint slopes only; $\mu,s,c$ are interpolated & charge-dependent generation spacing, as might arise in channel-dependent matching; empirical low-scale relation\\
$M_{uL}$ & $A_s\in\mathrm{span}(P_u,Y_L)$; $I_s,B_s$ free & discrete curvature containing the middle generation & common left-chiral response plus an up-sector channel; gives the viable matching rule $A_\nu=A_\ell$\\
$M_{dG}$ & $A_s\in\mathrm{span}(P_d,Y_L^2+Y_R^2)$ & a different charged curvature relation & quadratic charge strength plus a down-sector channel; resembles a radiative/Casimir-like dependence but is not derived from one\\
$M_H$ & $A_s=A_0+A_Y|Y_{R,s}|$ & affine right-charge dependence of curvature & minimal charge--curvature test; good charged interpolation, but its literal neutral continuation is kinematically impossible\\
$M_{B+uL}$ & both the $M_B$ slope and $M_{uL}$ curvature restrictions; $I_s$ free & two independent charged relations; rank seven & economical combined description selected after the scan; only its curvature rule has a viable neutral continuation\\
$M_{B+H}$ & both the $M_B$ slope and affine $|Y_R|$ curvature restrictions; $I_s$ free & two charged relations; rank seven; 9/9 LOO coverage & accurate charged control, but the literal $M_H$ neutral continuation has no physical root\\
$M_{B+dLR}$ & $M_B$ and $A_s=aP_{d,s}+bY_{L,s}Y_{R,s}$; $I_s$ free & rank seven; $2A_\ell-9A_u=0$ additionally tests $\mu$ and $c$, while $s$ remains interpolated & best fixed-$M_Z$ curvature-plane descendant; mixed left--right feature plus a free down channel; predicts $A_\nu=0$ under Dirac-like bookkeeping\\
$M_{\ell T}$ & $I_s,B_s$ free; $A_s=a(P_{u,s}-3Y_{L,s})=a(2P_{\ell,s}+T_{3L,s})$ & rank seven; $A_\ell:A_d:A_u=3:-1:1$; 9/9 LOO coverage & one common curvature direction combining a sector projector with left hypercharge or weak isospin; the two forms are identical only on charged sectors\\
$M_{B+\ell T}$ & simultaneous $M_B$ and $M_{\ell T}$ constraints; $I_s$ free & rank six; three charged relations; all nine masses identified & most stable complete model by LOO; its neutrino continuation depends on which charged-equivalent operator is taken as fundamental\\
$M_{B+\ell T_n}$ & $M_B$ plus the fixed integer curvature ray $(3n+2):-(n+1):n$ & rank six for every fixed integer $n$; all nine masses identified & finite constituent-channel hypothesis; $n=2$ is selection-aware conservative, while $n=3$ minimizes the central worst error\\
$M_{B+dLR}^{s14}$ & $M_{B+dLR}$ together with $A_d=B_d/14$ & rank six; turns $s$ into a genuine LOO prediction without changing $A_\nu=0$ & targeted cross-block matching; interpreted below as a possible RG-evolved $1/15$ boundary factor, conditional on an independently predicted matching scale\\
\shortstack{$M_{B+dG}$\\$^{[P_d-2G]}$} & $M_B$ plus the fixed curvature ray $A_s=a(P_{d,s}-2G_s)$, $G_s=Y_{L,s}^2+Y_{R,s}^2$ & rank six; all nine masses identified; $A_\nu=A_\ell/5$ & gauge-quadratic complete control retained to test a higher IO neutrino sum; not a global favorite\\
$M_{B,LR}$ & $M_B$ and $A_s=aY_{L,s}Y_{R,s}$; $I_s$ free & rank six; two curvature relations connect all three middle states & most accurate tested curvature ray, but weaker leave-one-out performance; a sign-sensitive left--right hypothesis rather than a standard gauge Casimir\\
\shortstack{$M_{2+B+}$\\$dLR$} & $M_{B+dLR}$ plus $I_\ell=I_d$ & rank six; slope, curvature, and centered-level relations & numerical blockwise rank-six winner; inherits the residual-selected and scale-sensitive M2 offset rule\\
$M_{2\nu}$ & charged M2 plus $A_\nu=A_\ell$, with $I_\nu,B_\nu$ free & the cross-sector curvature sum rule & phenomenological charged--neutral operator matching; it does not predict PMNS mixing or a microscopic mass mechanism\\
\bottomrule
\end{longtable}
\endgroup

Table~\ref{tab:modelmap} also fixes the historical record of the investigation.  M1, M2, the isolated rank-eight planes, $M_H$, $M_{dG}$, $M_{B,LR}$, and the unsuccessful auxiliary descendants are kept so that the path from the saturated basis to the surviving six-parameter hypotheses remains reproducible.  Their numerical diagnostics continue to be reported in compact comparison tables and appendices, but no independent microscopic narrative is built around each of them.  The physical discussion below is reserved for the complete $M_{B+\ell T_n}$ family, the targeted $s$ relation, and the repaired $dLR$ branch.

Three physical ideas are tested by that retained genealogy, and they should not be conflated.  First, the smooth coefficient $B_s$ may encode exponential hierarchies generated by channel-dependent matching or dimensional transmutation.  Second, $A_s$ measures the finite three-level curvature and may represent a sector-dependent shell correction or another non-smooth contribution.  Third, the feature planes ask whether those coefficients track simple chiral charges or sector projectors.  Linear charge dependence is reminiscent of charge-controlled suppression mechanisms, while $Y_L^2+Y_R^2$ resembles the positive charge combinations that can occur in gauge-loop strengths.  The product $Y_LY_R$ has a different status: it is sign sensitive and may serve as a diagnostic of dynamics that couples left- and right-chiral channels, but it is not the usual positive gauge Casimir and is not generated here from a specified diagram.  These are motivations only.  Standard-Model hypercharges do not explain why three generations exist, $|Y_R|$ is not itself a derived microscopic operator, and $P_u$ or $P_d$ distinguishes a sector by construction.  Thus the historical models document low-energy regularities and failed alternatives; they do not separately constitute candidate fundamental theories.

\subsection{Definitions of the leading projector models}\label{sec:leadingmodels}
The leading charged model is the slope restriction
\begin{equation}
\boxed{M_B:\quad B_s=b_R|Y_{R,s}|+b_LY_{L,s}.}
\label{eq:MB}
\end{equation}
It implies
\begin{equation}
B_\ell+9B_d-6B_u=0,
\qquad
\boxed{\RBu\equiv\frac{y_e y_d^9y_t^6}{y_u^6y_\tau y_b^9}=1.}
\label{eq:RB}
\end{equation}
The origin of the integer coefficients can be seen without a matrix calculation.  On the three charged sectors Eq.~\eqref{eq:MB} reads
\begin{equation}
B_\ell=b_R-\frac{b_L}{2},\qquad
B_d=\frac{b_R}{3}+\frac{b_L}{6},\qquad
B_u=\frac{2b_R}{3}+\frac{b_L}{6}.
\label{eq:MBcomponents}
\end{equation}
Eliminating $b_R$ and $b_L$ gives $B_\ell+9B_d-6B_u=0$.  Because
\begin{equation}
z_{s,1}=I_s-B_s+A_s,\qquad z_{s,2}=I_s-A_s,\qquad
z_{s,3}=I_s+B_s+A_s,
\label{eq:MBtriplet}
\end{equation}
the slope is $B_s=(z_{s,3}-z_{s,1})/2$ and contains no middle-generation value.  This statement is sometimes misunderstood.  It does \emph{not} mean that $M_B$ fails to reproduce $\mu$, $s$, and $c$.  The unrestricted coefficients
\begin{equation}
I_s=\frac{z_{s,1}+2z_{s,2}+z_{s,3}}{4},\qquad
A_s=\frac{z_{s,1}-2z_{s,2}+z_{s,3}}{4}
\end{equation}
adjust to any measured $z_{s,2}$, so the three middle-generation inputs are fitted exactly.  Equivalently, the unique null vector
\begin{equation}
\bm w_B=(1,9,-6,\,0,0,0,\,-1,-9,6)^{\trans}
\label{eq:wBexplicit}
\end{equation}
has zero entries in the middle-generation positions.  Thus $M_B$ genuinely tests an accurate cross-sector endpoint-slope relation, while its predictive content does not test the middle generation or the shell-like curvature.  Since $A_s$ remains unrestricted, $M_B$ alone also leaves $A_\nu$ and the absolute neutrino scale free.

Physically, $B_s=\tfrac12\ln(y_{s,3}/y_{s,1})$ measures the logarithmic separation of the endpoint generations.  The $M_B$ hypothesis says that this separation varies across charged sectors as a linear combination of the magnitudes of right-handed and signed left-handed hypercharges.  Such behavior could emerge from charge-dependent suppression factors, anomalous dimensions, or constituent-channel matching.  The fit does not distinguish these mechanisms, however, and the use of $|Y_R|$ rather than a derived operator makes the relation an empirical low-scale regularity until a microscopic construction reproduces it.

The best model whose literal neutral continuation is physical for both orderings is
\begin{equation}
\boxed{M_{uL}:\quad A_s=aP_{u,s}+bY_{L,s}.}
\label{eq:MuL}
\end{equation}
For charged sectors it gives $A_\ell+3A_d=0$ and
\begin{equation}
\boxed{\RuL\equiv
\frac{y_e y_d^3y_\tau y_b^3}{y_\mu^2y_s^6}=1.}
\label{eq:RuL}
\end{equation}
For a neutral state with $P_u=0$ and $Y_L=-1/2$,
\begin{equation}
A_\nu=-\frac b2=A_\ell.
\label{eq:MuLneutral}
\end{equation}
Thus its neutral rule is fixed without adding a parameter and coincides with the shared-curvature extension of M2.
The content of this formula is simple.  The projector $P_u=C(1+T)/2$ allows the up-quark curvature to have an additional contribution.  The remaining term is proportional to the left-handed hypercharge: $Y_L=-1/2$ for charged leptons and neutrinos, but $Y_L=1/6$ for both quark doublet components.  Consequently $M_{uL}$ requires $A_\ell=-3A_d$ while leaving $A_u$ independently adjustable, and it gives $A_\nu=A_\ell$ because the neutral and charged-lepton feature rows coincide.  This is a compact phenomenological rule; it is not yet an operator-matching derivation.

Because $A_s=\tfrac14\ln(y_{s,1}y_{s,3}/y_{s,2}^2)$, $M_{uL}$ specifically describes how far the middle state is displaced from a geometric progression.  A possible finite-system interpretation is a common left-chiral shell response for lepton and quark doublets plus an additional up-channel correction.  A possible field-theory interpretation is a sector-dependent matching or radiative contribution.  The projector $P_u$ nevertheless inserts the up-sector distinction by hand, so neither interpretation is established by the fit.

A genuinely different gauge-quadratic alternative is
\begin{equation}
\boxed{M_{dG}:\quad A_s=aP_{d,s}+bG_s,
\qquad G_s=Y_{L,s}^2+Y_{R,s}^2.}
\label{eq:MdG}
\end{equation}
It implies $17A_\ell-45A_u=0$.  Since $G_\ell=5/4$ and $G_\nu=1/4$,
\begin{equation}
A_\nu=\frac{A_\ell}{5},\qquad Q_\nu=Q_\ell^{1/5}.
\label{eq:MdGneutral}
\end{equation}
This is a distinct neutral prediction, although its charged constraint and extrapolation are less stable than those of $M_{uL}$.

The combination $G_s=Y_{L,s}^2+Y_{R,s}^2$ resembles the positive quadratic charge factors that can weigh gauge or radiative effects, while $P_d$ supplies a separate down-sector channel.  Thus $M_{dG}$ tests whether the discrete curvature behaves as a quadratic charge response rather than the linear left-charge response of $M_{uL}$.  Its pairwise continuation ambiguity shows that this physical reading is substantially less stable than the compact formula may suggest.

\subsection{Previous affine hypercharge-curvature test \texorpdfstring{$M_H$}{MH}}\label{sec:MHdef}
We use the convention $Q=T_3+Y$ and define the sector operator
\begin{equation}
\widehat H=|Y_R|,\qquad (H_\ell,H_d,H_u)=\left(1,\frac13,\frac23\right).
\end{equation}
Its spectral projectors are
\begin{equation}
P_s(\widehat H)=\prod_{s'\ne s}\frac{\widehat H-H_{s'}}{H_s-H_{s'}},
\qquad P_sP_{s'}=\delta_{ss'}P_s,\qquad \sum_sP_s=1.
\label{eq:Hprojectors}
\end{equation}
A completely general curvature $A_\ell P_\ell+A_dP_d+A_uP_u$ merely reproduces the saturated nine-parameter basis in Eq.~\eqref{eq:charged_sat}.  To retain predictive content, $M_H$ imposes the minimal affine dependence
\begin{equation}
\ln y_{H,s,g}=I_s+B_s(g-2)+\bigl(A_0+A_YH_s\bigr)h_g.
\label{eq:MH}
\end{equation}
The model has eight parameters and rank eight.  Its single left-null vector is
\begin{equation}
\bm w_H=(-1,-1,+2,+2,+2,-4,-1,-1,+2)^{\trans},
\end{equation}
which is equivalent to the curvature relation $2A_u=A_\ell+A_d$ and the exact invariant
\begin{equation}
\boxed{\Rhyper(\mu)\equiv
\frac{y_u^2y_t^2y_\mu^2y_s^2}{y_c^4y_ey_\tau y_dy_b}=1.}
\label{eq:RH}
\end{equation}
The choice of $|Y_R|$ and the affine restriction were examined after the charged data and are therefore post-selected hypotheses, not symmetry-derived predictions.

The physical question posed by $M_H$ is intentionally minimal: can all three charged curvatures lie on one affine line when ordered by the magnitude of right-handed hypercharge?  The charged data answer approximately yes, but the sterile-neutrino endpoint $Y_R=0$ extrapolates far outside the oscillation-allowed range.  This failure is useful because it separates a visually successful charged interpolation from a physically admissible cross-sector law.

\subsection{Neutral extension of M2}\label{sec:M2nudef}
A fully independent neutral triplet,
\begin{equation}
z_{\nu i}=I_\nu+B_\nu(i-2)+A_\nu h_i,
\end{equation}
has three coefficients for three masses and no prediction.  The constrained extension $M_{2\nu}$ instead fixes the neutral curvature to the charged-lepton curvature appearing in M2:
\begin{equation}
z_{\nu i}=I_\nu+B_\nu(i-2)+A_\ell h_i,\qquad i=1,2,3.
\label{eq:M2nu}
\end{equation}
For Dirac neutrinos one may take $z_{\nu i}=\ln y_{\nu i}$.  More generally, $z_{\nu i}=\ln(m_i/m_*)$, where the arbitrary reference mass cancels from all relations.  The index $i$ labels mass eigenstates, not flavor states; these ans\"atze do not predict the PMNS matrix.

For the state order in Eq.~\eqref{eq:ordering} followed by $(\nu_1,\nu_2,\nu_3)$, the $12\times10$ design matrix has rank ten and a two-dimensional left-null space.  One vector reproduces Eq.~\eqref{eq:R2}; the second gives
\begin{equation}
\ln\frac{m_1m_3}{m_2^2}=\ln\frac{m_em_\tau}{m_\mu^2},
\end{equation}
or
\begin{equation}
\boxed{\Rthree\equiv\frac{m_1m_3m_\mu^2}{m_2^2m_em_\tau}=1,\qquad
Q_\nu\equiv\frac{m_1m_3}{m_2^2}=e^{4A_\ell}.}
\label{eq:R3}
\end{equation}
Allowing $A_\nu$ to vary independently would add an eleventh coefficient and remove Eq.~\eqref{eq:R3}.  The shared curvature is therefore the entire predictive assumption of $M_{2\nu}$.  Equation~\eqref{eq:R3} must be evaluated with the direct charged-lepton invariant $Q_\ell=y_ey_\tau/y_\mu^2$; substituting the globally fitted M2 value of $A_\ell$ would instead be a conditional least-squares projection.  The full $M_{2\nu}$ model also inherits the imperfect charged relation $R_2=1$.  The projector model $M_{uL}$ produces the same cross-sector rule automatically, while replacing $R_2=1$ by the more accurate charged constraint in Eq.~\eqref{eq:RuL}.

Thus $M_{2\nu}$ is not a new microscopic neutrino model.  It is the statement that whatever matching maps the charged-lepton and neutral mass operators preserves one particular dimensionless second difference in log mass.  It fixes an absolute eigenvalue scale only after the two measured oscillation splittings are supplied; it says nothing about the PMNS matrix, Majorana phases, or the ultraviolet origin of the neutral operator.

\subsection{Inputs and statistical convention}\label{sec:data}
The primary charged inputs are SM Yukawa eigenvalues in the $\MSbar$ scheme, evolved to $M_Z$ using the PDG-2024 profile of Antusch, Hinze, and Saad \cite{PDG2024,Antusch2026}.  The values entering the invariants are
\begin{align}
y_u&=(7.04\pm0.15)\times10^{-6}, & y_c&=(3.56\pm0.06)\times10^{-3},\\
y_t&=0.967\pm0.004, & y_e&=(2.77713\pm0.00036)\times10^{-6},\\
y_\mu&=(5.85042\pm0.00075)\times10^{-4}, & y_\tau&=(0.99378\pm0.00014)\times10^{-2},\\
y_d&=(1.540\pm0.020)\times10^{-5}, & y_s&=(3.06\pm0.04)\times10^{-4},\\
y_b&=(1.630\pm0.009)\times10^{-2}.
\label{eq:inputs}
\end{align}
Using one scheme and one scale is essential: pole lepton masses, light-quark masses quoted at $2\,\GeV$, charm and bottom masses at their own scales, and the top mass cannot be inserted directly into one putative ultraviolet flavor relation.

Reference~\cite{Antusch2026} provides marginal one-standard-deviation intervals but not a full covariance matrix across flavors and scales.  We therefore treat propagated input errors as diagnostics and introduce an illustrative relative model discrepancy $\sth$ through
\begin{equation}
\sigma_i^2=\ln^2\left(1+\frac{\delta y_i}{y_i}\right)+\ln^2(1+\sth).
\label{eq:sigma}
\end{equation}
For diagonal $W=\diag(\sigma_i^{-2})$,
\begin{equation}
\chi_k^2(\bm\theta_k)=(\bm y_{\log}-X_k\bm\theta_k)^{\trans}W(\bm y_{\log}-X_k\bm\theta_k),
\qquad k=2,B,uL,dG,H.
\end{equation}
Since each model has a one-dimensional left-null space, the exact minimum is
\begin{equation}
\chi^2_{k,\min}=\frac{\ln^2R_k}{\sigma^2_{\ln R_k}},\qquad \nu=9-8=1,
\label{eq:exactchi}
\end{equation}
where the coefficients of $\bm w_k$ determine $\sigma^2_{\ln R_k}$.  These $\chi^2$ values are conditional on $\sth$; the exact parameter-elimination ratios are not.

For comparisons that do not depend on the arbitrary numerical choice of $\sth$, we also report the Euclidean distance from the measured log-spectrum to the model hyperplane,
\begin{equation}
 \boxed{\dperp^{(k)}=\frac{|\bm w_k^{\trans}\bm y_{\log}|}{\|\bm w_k\|_2}},\qquad
 \epsilon_{{\rm rms},k}=\frac{\dperp^{(k)}}{\sqrt{9}}.
\label{eq:dperp}
\end{equation}
The normalization removes the arbitrary rescaling of a null vector.  The second quantity is the minimum root-mean-square log displacement required to place all nine points on the corresponding rank-eight hyperplane under equal weights.  It is a geometric descriptive measure, not a probability or a replacement for a specified covariance model.

For a model of rank $r<8$, let the columns of $N$ be an orthonormal basis of its $(9-r)$-dimensional left-null space.  We then use the direct generalization
\begin{equation}
d_{\rm subspace}=\|N^{\trans}\bm y_{\log}\|_2.
\label{eq:dsubspace}
\end{equation}
This counts all independent departures without privileging a particular normalization or ordering of the sum rules.

Finally, the last column of Table~\ref{tab:leadingcharged} uses
\begin{equation}
 \Delta_{\max}=100\max_i\left|\frac{y_{i,{\rm data}}}{y_{i,{\rm fit}}}-1\right|.
\label{eq:maxshiftdefinition}
\end{equation}
Writing the orientation explicitly matters because $y_{\rm data}/y_{\rm fit}-1$ and $y_{\rm fit}/y_{\rm data}-1$ are not identical for finite shifts.  The numerical audit also evaluates the symmetric alternative $100\max_i|\Delta\ln y_i|$.

An in-sample conditional shift is not yet a prediction.  We therefore also remove each mass in turn, refit the fixed feature subspace to the other eight inputs, and predict the omitted value.  If deleting a state lowers the design rank, that mass is reported as \emph{unidentified} rather than assigned a spurious leave-one-out (LOO) error.  LOO tests the interpolation of a predeclared subspace; it does not undo the post-selection involved in choosing that subspace from the same spectrum.

As a covariance sensitivity test, we also use
\begin{equation}
\Sigma_{\rm th}(\rho)=\ln^2(1+\sth)[(1-\rho)\bm1+\rho B],\qquad0\le\rho\le1,
\end{equation}
where $B_{ij}=1$ for states in the same sector, or alternatively the same generation, and zero otherwise.  These are scenarios, not estimates of the unknown physical covariance.

For neutrinos we use the NuFIT~6.1 central splittings, including the tabulated atmospheric likelihoods \cite{NuFIT2025}
\begin{equation}
\Delta m_{21}^2=7.537\times10^{-5}\,\eV^2,
\end{equation}
\begin{equation}
\Delta m_{31}^2=+2.511\times10^{-3}\,\eV^2\quad(\mathrm{NO}),\qquad
\Delta m_{32}^2=-2.483\times10^{-3}\,\eV^2\quad(\mathrm{IO}).
\label{eq:nufit}
\end{equation}
These are two independent measurements, not three absolute masses.  Consequently, we do not quote a new aggregate $\chi^2/\nu$ for any neutral extension.

\section{Charged-sector results}\label{sec:chargedresults}
\subsection{M2 baseline and leading projector models at \texorpdfstring{$M_Z$}{MZ}}
For the baseline M2 model, the common-scale data give
\begin{equation}
\Rtwo(M_Z)=0.401892,\qquad\ln\Rtwo=-0.9114,
\end{equation}
and
\begin{equation}
\chi^2_{2,\min}=1.3887\qquad(\sth=0.25).
\end{equation}
The projector scan yields substantially smaller conditional residuals without changing the rank.  Table~\ref{tab:leadingcharged} compares M2 with the leading models.  The experimental-only column is shown as a diagnostic, not as a calibrated significance, because correlations and model discrepancy are unknown.

\begin{table}[H]
\centering
\caption{M2 and leading projector models.  Each charged model has eight parameters, rank eight, and one exact charged constraint.}
\label{tab:leadingcharged}
\small
\begin{tabular}{lrrrrr}
\toprule
model & block & $\chi^2(\sth=0.25)$ & $\dperp$ & $\chi^2_{\rm exp}$ & $\Delta_{\max}$ (\%)\\
\midrule
M2       & $I_s$ & 1.3887 & 0.26315 & 953.39 & 16.44\\
$M_{dG}$ & $A_s$ & 0.3170 & 0.12592 & 68.85  & 9.18\\
$M_{uL}$ & $A_s$ & 0.0648 & 0.05687 & 24.74  & 4.31\\
$M_H$    & $A_s$ & 0.0463 & 0.04813 & 11.60  & 3.16\\
$M_B$    & $B_s$ & 0.00119& 0.00771 & 0.431  & 0.45\\
\bottomrule
\end{tabular}
\end{table}

\begin{figure}[H]
\centering
\includegraphics[width=0.98\textwidth]{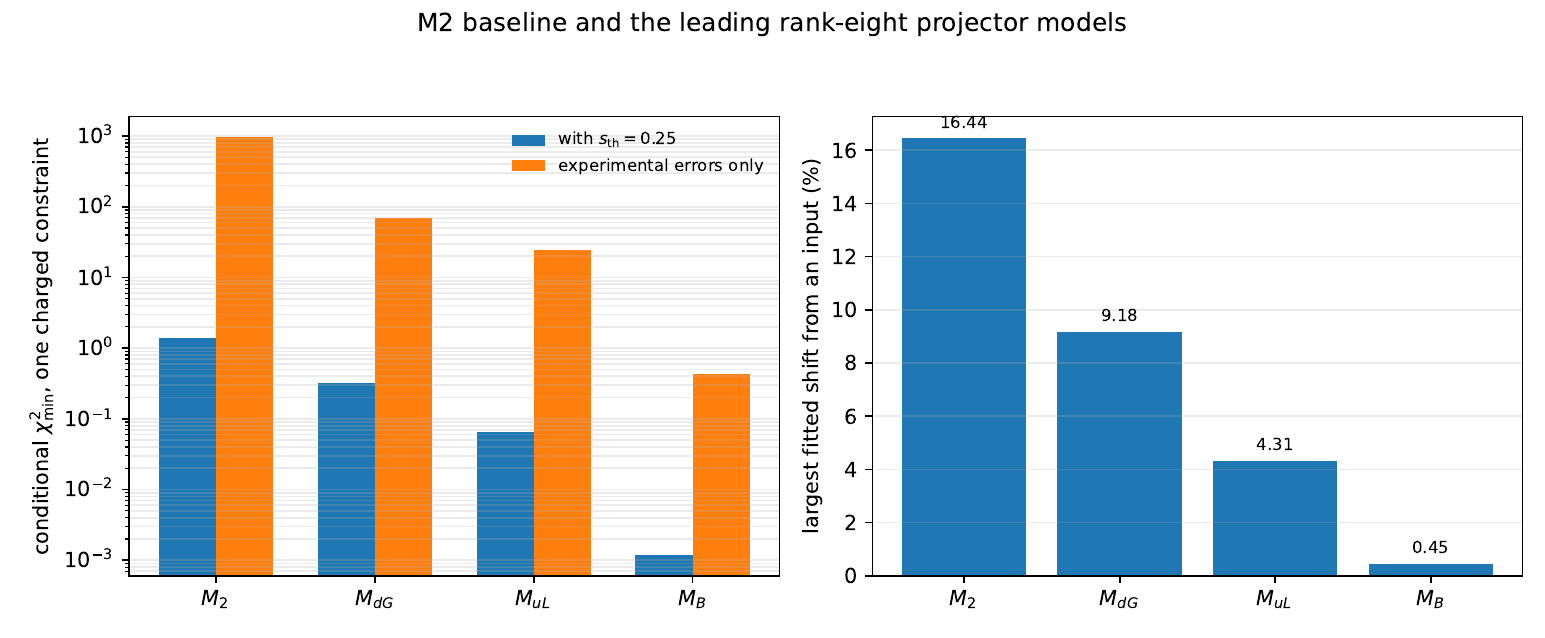}
\caption{Direct comparison of the M2 baseline with the leading rank-eight projector models.  Left: conditional minima with the illustrative structural discrepancy and with marginal experimental errors alone.  Right: the explicitly oriented discrepancy $100\max_i|y_{i,{\rm data}}/y_{i,{\rm fit}}-1|$.  The ordering is post-selected and is not a model-selection probability.}
\label{fig:projectorcomparison}
\end{figure}

\subsection{Interpretation of the charged ranking}
The smallest fixed-$M_Z$ charged residual belongs to $M_B$: the data give $R_B=1.12571$, $\chi^2_{\min}=0.00119$ for $\sth=0.25$, $\chi^2_{\rm exp}=0.431$, and $d_\perp=0.00771$.  This is the closest charged-sector hyperplane found in the declared feature dictionary and should be retained as a genuine empirical result.  In the conditional fit the shifts of $\mu$, $s$, and $c$ vanish to numerical precision, while the largest endpoint shift is only $0.453\%$.  The limitation is not a failure to reproduce the second generation; it is that those three values do not participate in the sole cross-sector test.  Nor does the relation constrain $A_\nu$.

The model $M_{uL}$ gives $R_{uL}=1.55349$, $\chi^2_{\min}=0.0648$, and $d_\perp=0.05687$.  Its charged fit is slightly less close than that of $M_H$, but its neutral continuation is physical and parameter free.  More importantly for the shell interpretation, it restricts $A_s$ itself and uses all three generations in the charged-lepton and down-quark sectors.  The distinction is essential: a superior charged interpolation need not test the proposed curvature or supply a viable neutral prediction.  The model $M_{dG}$ is retained not because it wins the charged ranking, but because it provides a genuinely different neutrino curvature and thereby exposes the continuation uncertainty.

\subsection{Nested predictive audit of \texorpdfstring{$M_B$}{MB}: seven and six parameters}\label{sec:rankseven}
The parameter count is clearest when a model is labeled by the sector-space dimensions $(d_I,d_B,d_A)$ of its offset, slope, and curvature blocks.  The saturated reference is $(3,3,3)$ and $M_B$ is $(3,2,3)$: it has removed one slope direction but no curvature direction.  To make the middle generation participate while preserving the successful $M_B$ relation and three independent normalizations, the natural next steps are therefore $(3,2,2)$ at rank seven and $(3,2,1)$ at rank six.

Among all 52 curvature planes in the $(3,2,2)$ class, the smallest fixed-$M_Z$ residual is obtained for
\begin{equation}
\boxed{M_{B+dLR}:\quad B_s\in\mathrm{span}(|Y_R|,Y_L),\qquad
A_s=aP_{d,s}+bY_{L,s}Y_{R,s},}
\label{eq:MBdLR}
\end{equation}
with independent $I_s$.  Since $(Y_LY_R)_{\ell,d,u}=(1/2,-1/18,1/9)$, eliminating $a,b$ gives
\begin{equation}
2A_\ell-9A_u=0,\qquad
\boxed{R_{dLR}\equiv\frac{Q_\ell^2}{Q_u^9}=1},\qquad
Q_s\equiv\frac{y_{s,1}y_{s,3}}{y_{s,2}^2}=e^{4A_s}.
\label{eq:RdLR}
\end{equation}
The down projector absorbs $A_d$, so this new relation contains $\mu$ and $c$ but not $s$.  Thus the model has two genuine charged constraints, yet it does not predict every middle-generation mass.  Its conditional metrics are
\begin{equation}
\chi^2_{\min}=0.01336,\qquad d_{\rm subspace}=0.02586,
\qquad\Delta_{\max}=1.99\%.
\label{eq:MBdLRmetrics}
\end{equation}
The mixed feature $Y_LY_R$ is physically suggestive only in a limited sense: it changes sign between the charged-lepton, down, and up sectors and could diagnose a mechanism coupling both chiral channels.  Unlike $Y_L^2+Y_R^2$, it is not a positive gauge Casimir.  The separate $P_d$ term is an explicitly inserted down-channel freedom.

The one-dimensional curvature reduction tests all three sectors.  Among 14 distinct feature rays, the best is
\begin{equation}
\boxed{M_{B,LR}:\quad B_s\in\mathrm{span}(|Y_R|,Y_L),\qquad
A_s=aY_{L,s}Y_{R,s},}
\label{eq:MBLR}
\end{equation}
so that $A_\ell:A_d:A_u=9:-1:2$ and
\begin{equation}
A_\ell+9A_d=0,\qquad A_u+2A_d=0.
\label{eq:MBLRrelations}
\end{equation}
This six-parameter model has three charged constraints and every $A_s$ is tied to the same coefficient.  It remains inside the illustrative 25\% in-sample tolerance,
\begin{equation}
\chi^2_{\min}=1.65243,\qquad d_{\rm subspace}=0.28726,
\qquad\Delta_{\max}=20.73\%,
\label{eq:MBLRmetrics}
\end{equation}
but its LOO behavior is poor: the largest omitted-mass error is $194\%$, and the largest second-generation error is $52.7\%$ for $\mu$.  Hence rank six creates more formal predictions but not automatically more reliable ones.

The full nested audit also tested 52 offset-plane reductions, two slope rays contained in the $M_B$ plane, all $52^2=2704$ simultaneous offset--curvature plane pairs, 208 plane-plus-slope-ray combinations, 28 offset or curvature rays, and two shared-$Y_L$ cross-block ties.  The numerical blockwise rank-six winner is $M_{2+B+dLR}$: it adds $I_\ell=I_d$ to Eq.~\eqref{eq:MBdLR} and gives $\chi^2=1.39235$, $\Delta_{\max}=16.40\%$, and a worst LOO error of $57.4\%$.  It is not preferred physically because the offset rule was motivated by the M1 residuals and is strongly scale sensitive.  A focused follow-up then asks whether the poor rule $I_\ell=I_d$ can be replaced by more structured six-parameter relations while keeping the same $M_B$ slope plane and the same $dLR$ curvature plane.  Three such repairs---$I_\ell-I_d=A_\ell/3$, $I_\ell-I_d+(B_\ell-B_d)/3=0$, and $I_\ell-I_d+B_d/16=0$---prove much more successful and are discussed below.  Reducing the $M_B$ slope plane itself to one ray is disastrous ($\chi^2\ge380$), and sharing the normalized $Y_L$ coefficient between the slope and curvature blocks gives $\chi^2=149$; both options are rejected by the charged data.

\begin{table}[H]
\centering
\caption{Predictive audit of representative descendants of $M_B$.  ``Free'' in the LOO column means that deleting the stated mass lowers the design rank, so no unique prediction exists.}
\label{tab:MBdescendants}
\footnotesize
\begin{tabularx}{\textwidth}{@{}lrrrrr>{\raggedright\arraybackslash}X@{}}
\toprule
model & rank & constraints & $d_{\rm subspace}$ & $\chi^2_{25\%}$ & $\Delta_{\max}$ & largest LOO error / unidentified states\\
\midrule
$M_B$ & 8 & 1 & 0.00771 & 0.00119 & 0.45\% & 12.6\% among endpoints; $\mu,s,c$ free\\
$M_{B+dLR}$ & 7 & 2 & 0.02586 & 0.01336 & 1.99\% & 24.9\%; $s$ free\\
$M_{B+H}$ & 7 & 2 & 0.04875 & 0.04755 & 3.16\% & 23.4\%; none free\\
$M_{B+uL}$ & 7 & 2 & 0.05739 & 0.06596 & 4.31\% & 45.5\%; $c$ free\\
$M_{2+B+dLR}$ & 6 & 3 & 0.26350 & 1.39235 & 16.40\% & 57.4\%; none free\\
$M_{2+B+dLR}^{[A_\ell/3]}$ & 6 & 3 & 0.02685 & 0.01920 & 1.86\% & 3.82\%; none free\\
$M_{2+B+dLR}^{[\delta=1/3]}$ & 6 & 3 & 0.03278 & 0.02862 & 1.91\% & 5.78\%; none free\\
$M_{2+B+dLR}^{[B_d/16]}$ & 6 & 3 & 0.02784 & 0.01547 & 1.95\% & 3.91\%; none free\\
$M_{B,LR}$ & 6 & 3 & 0.28726 & 1.65243 & 20.73\% & 193.7\%; none free\\
\bottomrule
\end{tabularx}
\end{table}

\paragraph{How to read Table~\ref{tab:MBdescendants}.}
The first five numerical columns describe a model fitted with all nine known masses.  Smaller $d_{\rm subspace}$, $\chi^2$, and $\Delta_{\max}$ mean that the model passes closer to the observed points.  The last column asks the harder question: what happens when one mass is hidden and reconstructed from the other eight?  ``$s$ free'' for $M_{B+dLR}$ does not mean a large numerical error; it means that no unique number exists.  The coefficient of $P_d$ can be changed to obtain any $m_s$ without violating the model's other two relations.  Thus the very small $1.99\%$ conditional displacement is not yet a strange-quark prediction.

Sampling the quoted experimental input errors 4000 times leaves $M_{B+dLR}$ as the best $(3,2,2)$ plane in $77.8\%$ of draws, $M_{B+H}$ in $20.7\%$, and $M_{B+uL}$ in $1.47\%$.  The ray $Y_LY_R$ wins all draws within the 14-member $(3,2,1)$ class.  These frequencies measure sensitivity to the quoted inputs only; they are not discovery probabilities and do not account for the chosen feature dictionary or the uncalibrated structural discrepancy.

The earlier $M_{B+uL}$ model remains important because its curvature continuation permits both neutrino orderings:
\begin{equation}
\boxed{M_{B+uL}:\quad
B_s=b_R|Y_{R,s}|+b_LY_{L,s},\qquad
A_s=aP_{u,s}+bY_{L,s}.}
\label{eq:MBuL}
\end{equation}
It has $\chi^2_{\min}=0.06596$, $d_{\rm subspace}=0.05739$, and $\Delta_{\max}=4.31\%$.  Only $A_\nu=A_\ell$ is continued neutrally.  A literal continuation of the $M_B$ slope gives $B_\nu=-3.1843$ and, together with $A_\nu=A_\ell$, would imply $m_1/m_2\simeq6.86$, contradicting the positive solar splitting.  Thus the slope continuation is falsified even though the curvature continuation remains viable.

\begin{figure}[H]
\centering
\includegraphics[width=0.98\textwidth]{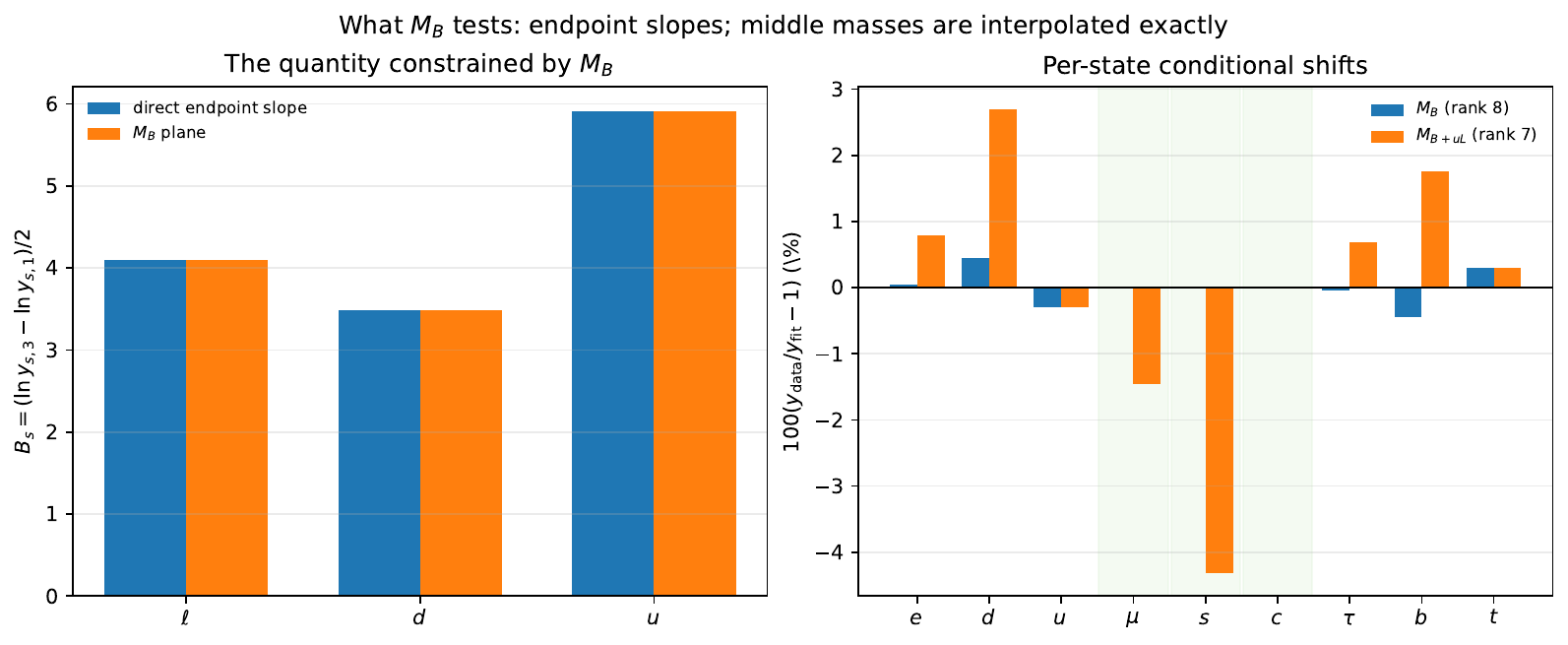}
\caption{What $M_B$ does and does not test.  Left: direct endpoint slopes and their projection onto the $M_B$ feature plane.  Right: per-state conditional shifts.  The shaded middle-generation entries are reproduced exactly by $M_B$ because $I_s$ and $A_s$ remain sector resolved.}
\label{fig:MBgeneration}
\end{figure}

\begin{figure}[H]
\centering
\includegraphics[width=0.99\textwidth]{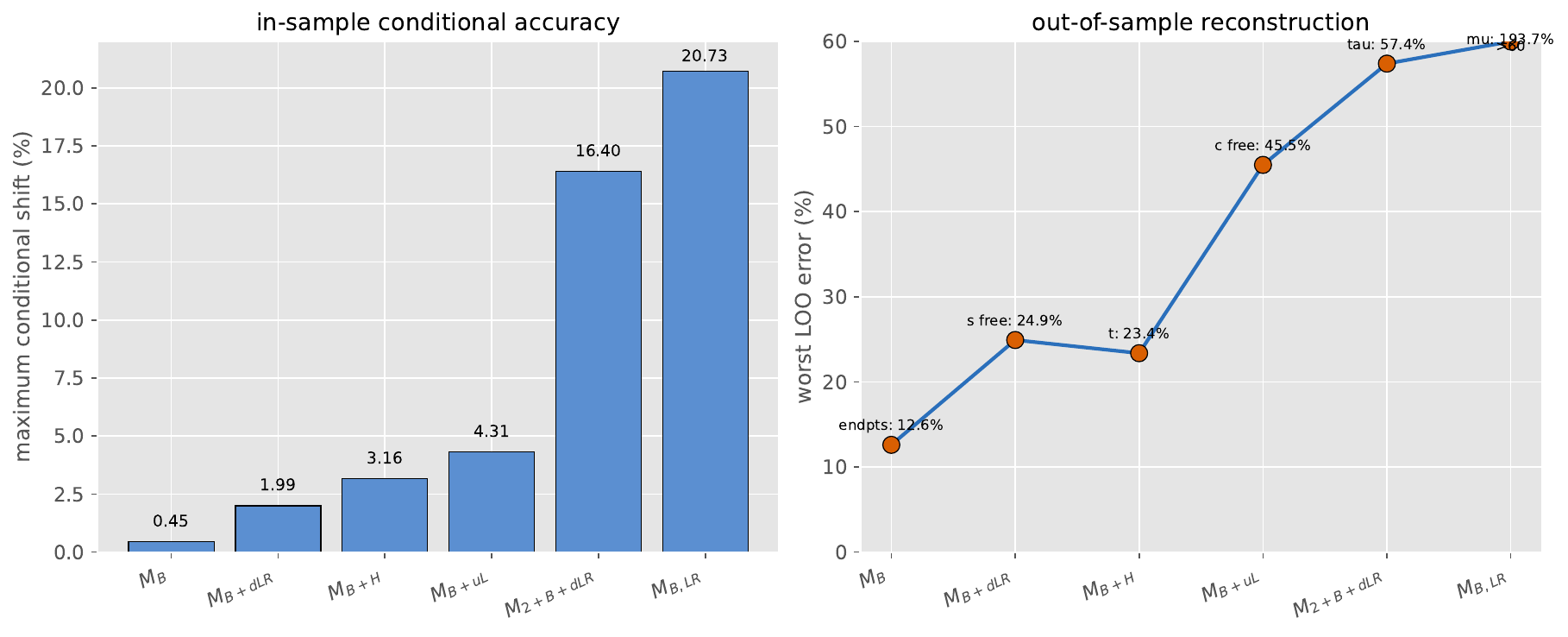}
\caption{In-sample accuracy and leave-one-out prediction are different diagnostics.  The seven-parameter $M_{B+dLR}$ gives the smallest conditional displacement, but $s$ is still unidentified.  The six-parameter models identify all omitted masses, at the cost of substantially larger prediction errors.  Each displayed subspace and error convention is defined in the text; the choice among subspaces remains exploratory.}
\label{fig:MBpredictiveaudit}
\end{figure}

\paragraph{How to read Fig.~\ref{fig:MBpredictiveaudit}.}
The figure has two panels, with the models listed horizontally in both.  The left panel gives the largest displacement of a known mass when all nine points are supplied: a shorter bar means a more accurate simultaneous description.  The right panel gives the worst leave-one-out error after one mass is hidden; a lower marker now means a more reliable genuine prediction.  The label beside a marker identifies the state responsible for the worst case, while a value printed above the plotting range denotes a very large error.  Thus $M_{B+dLR}$ looks excellent in the left panel (only $1.99\%$), but the right panel marks $s$ as unidentified: the model supplies no unique number for it.  The figure therefore separates two different properties---proximity to already known data and the ability to reconstruct a deliberately hidden mass.

\subsection{Post-selection and feature degeneracy}
The 156 rank-eight charged models are not 156 independent statistical trials.  Several feature pairs span the same two-dimensional plane on the three charged sectors, and their invariant is then identical.  They must nevertheless be kept distinct in a neutral audit when their values differ at $Y_R=0$, $Y_L=-1/2$, and $T_{3L}=+1/2$.  The nested audit adds 52 natural rank-seven curvature planes and 14 natural rank-six curvature rays, besides the control reductions listed above.  Thus the very small residual of $M_{B+dLR}$ and the 100\% input-resampling win frequency of the $Y_LY_R$ ray are descriptive within the declared dictionary, not corrected significances.  We use the full scan to expose exact relations, but do not attach a naive look-elsewhere probability to the minimum.  A valid confirmation requires either an out-of-sample observable or a microscopic derivation of the selected feature plane or ray.

\subsection{Covariance, centering, and scale diagnostics}
Positive correlations can either increase or decrease the variance of a left-null combination.  Reference~\cite{Antusch2026} supplies marginal errors, not the cross-flavor covariance needed to assign calibrated significances to M2 or the projector models.  This is why the exact ratios and the two discrepancy conventions in Table~\ref{tab:leadingcharged} are reported separately.

The centered coordinate $g-2$ in M2 is fixed a priori by the reflection-symmetric numbering of three generations.  This choice matters because the constraint $I_\ell=I_d$ is not invariant under a shift of the coordinate origin.  For a fixed general center $g_0$, the same construction implies
\begin{equation}
\Rtwo(g_0)\equiv
\left(\frac{y_e}{y_d}\right)^{5-2g_0}
\left(\frac{y_\mu}{y_s}\right)^2
\left(\frac{y_\tau}{y_b}\right)^{2g_0-3}=1,
\label{eq:R2g0}
\end{equation}
which reduces to Eq.~\eqref{eq:R2} only for $g_0=2$.  Alternative fixed centers $g_0=1,2,3$ give conditional minima $8.026$, $1.389$, and $1.666$.  If $g_0$ is continuously fitted, the optimum $g_0\simeq2.374$ merely selects the crossing point of the smooth charged-lepton and down-quark trends.  It is therefore a hidden ninth parameter and saturates the nine charged data; it must not be counted as a predictive eight-parameter fit.

The ratios are parameter-elimination invariants at fixed $\mu$, not RG invariants.  Using the SM running table of Ref.~\cite{Antusch2026}, $R_2$ crosses unity near $4.1\,\mathrm{TeV}$.  That crossing is an a posteriori feature: if the renormalization scale were adjusted to satisfy the only M2 constraint, the scale would be a hidden ninth fitting parameter and the model would be saturated.  The projector scan is therefore performed at the fixed common scale $M_Z$.

\begin{figure}[H]
\centering
\includegraphics[width=0.82\textwidth]{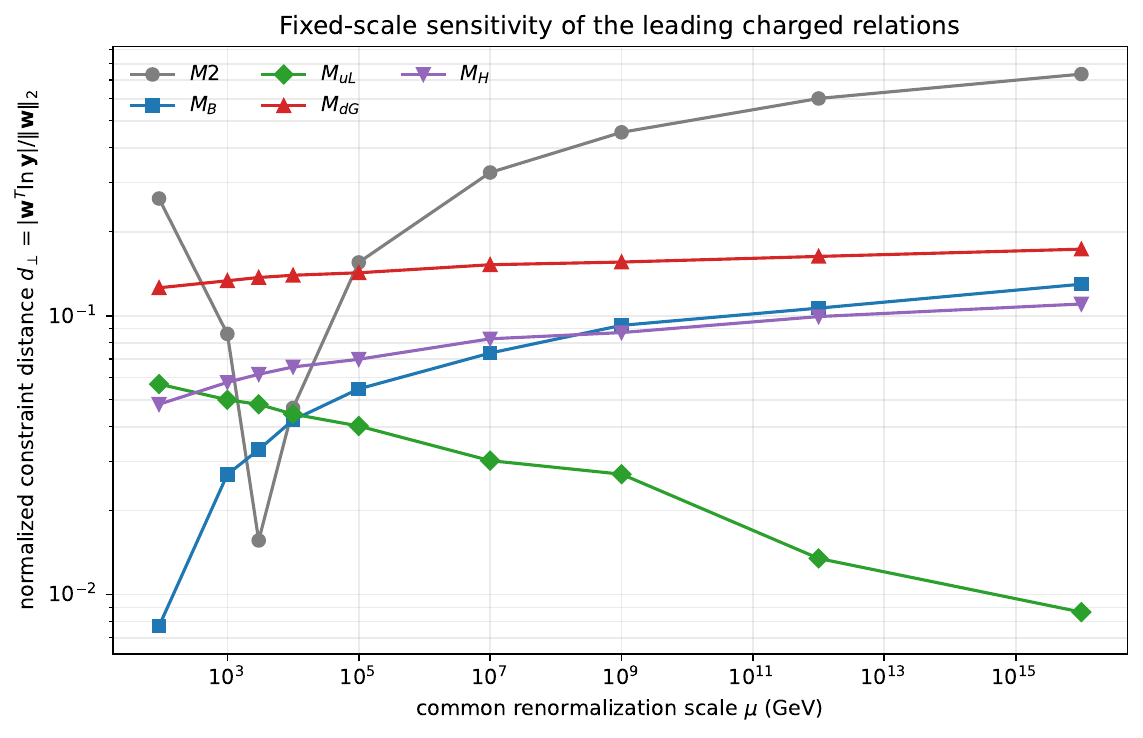}
\caption{Fixed-scale sensitivity of the leading charged relations.  The very small $M_B$ residual is specific to the low-scale endpoint relation and grows under SM running.  In contrast, the curvature relation of $M_{uL}$ becomes progressively closer over the tabulated range.  This supports treating $M_{uL}$ as the more relevant exploratory ultraviolet candidate, but it is not an RG-invariance proof and does not determine a matching scale.}
\label{fig:rgstability}
\end{figure}

\paragraph{Fixed-scale robustness across the tabulated running range.}
Although the scale must not be fitted to force a relation, it is informative to evaluate every already-defined invariant at the same set of externally tabulated scales.  The unnumbered profile table below and Fig.~\ref{fig:rgstability} use the normalized distance in Eq.~\eqref{eq:dperp}.  They are a sensitivity diagnostic, not nine independent experiments: all rows are connected by the SM renormalization-group equations.

\begin{table}[H]
\centering
\caption*{Normalized subspace distance at selected common scales from the running profile of Ref.~\cite{Antusch2026}.  For rank-eight rows it reduces to $\dperp$; for lower ranks Eq.~\eqref{eq:dsubspace} is used.  No scale is optimized.}
\begin{tabular}{lrrrr}
\toprule
model & $M_Z$ & $10^3\,\GeV$ & $10^7\,\GeV$ & $10^{16}\,\GeV$\\
\midrule
M2       & 0.26315 & 0.08607 & 0.32615 & 0.73492\\
$M_B$    & 0.00771 & 0.02697 & 0.07346 & 0.12972\\
$M_{uL}$ & 0.05687 & 0.04994 & 0.03022 & 0.00865\\
$M_{dG}$ & 0.12592 & 0.13345 & 0.15232 & 0.17323\\
$M_H$    & 0.04813 & 0.05768 & 0.08264 & 0.11018\\
$M_{B+dLR}$ & 0.02586 & 0.03179 & 0.07352 & 0.13205\\
$M_{B+uL}$  & 0.05739 & 0.05676 & 0.07944 & 0.13001\\
$M_{B,LR}$  & 0.28726 & 0.28083 & 0.26907 & 0.27109\\
$M_{2+B+dLR}$ & 0.26350 & 0.09105 & 0.33461 & 0.74639\\
\bottomrule
\end{tabular}
\end{table}

The scale diagnostic changes the physical emphasis without changing the fixed-$M_Z$ ranking.  The combined rank-seven distances of $M_{B+dLR}$ and $M_{B+uL}$ both approach about $0.13$ at $10^{16}\,\GeV$ because the growing $M_B$ slope violation eventually dominates.  The isolated $M_{uL}$ curvature relation is still the only leading neutral-capable curvature rule whose violation decreases steadily toward high scales.  The rank-six $M_{B,LR}$ distance is nearly scale independent but never small, while $M_{2+B+dLR}$ is close only around a few TeV and becomes the least stable descendant in the ultraviolet.  A future microscopic construction should predict its matching scale before comparison with the data; otherwise choosing the most favorable scale would again consume predictive content.

\section{Neutral-sector comparison}\label{sec:neutralresults}
\subsection{Neutral-sector convention and physical domain}
The neutral triplet is assigned its own offset and slope because a Dirac Yukawa coupling, the Weinberg operator, and a seesaw matching map need not share the charged-fermion normalization.  Only the discrete curvature is extrapolated.  We use
\begin{equation}
(Y_R,Y_L,T_{3L},B-L,P_C)_\nu=(0,-1/2,+1/2,-1,0).
\end{equation}
This is a Dirac-like bookkeeping convention: $Y_R=0$ refers to a sterile right-handed neutrino, while $P_\ell=1$ follows from defining $P_\ell=1-P_C$.  A Majorana mass generated by the Weinberg operator or by a seesaw is an operator-matching problem, not merely the same feature vector with another numerical mass.  Consequently, projector extensions that depend on $Y_R$ are mechanism dependent.  The relation $A_\nu=A_\ell$ in $M_{uL}$ is comparatively transparent because the neutral and charged-lepton rows coincide in the two features actually used, $(P_u,Y_L)=(0,-1/2)$, but even this equality remains a hypothesis about the matched mass eigenvalues.
For the NuFIT~6.1 splittings, the function $Q_\nu=m_1m_3/m_2^2$ satisfies $0<Q_\nu<1$ for inverted ordering and reaches a maximum $Q_\nu^{\max}=2.9303$ for normal ordering.  A proposed continuation outside these domains has no physical root.

\subsection{Normal-ordering prediction of the \texorpdfstring{$Y_LY_R$}{YL YR} descendants}
For the bookkeeping row of a sterile right-handed neutrino, $Y_{R,\nu}=0$ and $P_{d,\nu}=0$.  Both Eq.~\eqref{eq:MBdLR} and Eq.~\eqref{eq:MBLR} therefore imply
\begin{equation}
\boxed{A_\nu=0,\qquad Q_\nu=e^{4A_\nu}=1.}
\label{eq:Qnuone}
\end{equation}
This result does not use a fitted charged coefficient: it follows from the neutral feature row itself.  With the central splittings in Eq.~\eqref{eq:nufit}, the finite positive solution exists only for normal ordering,
\begin{equation}
(m_1,m_2,m_3)=(0.001551,\ 0.008819,\ 0.050134)\,\eV,
\qquad \boxed{\sum m_\nu=0.060504\,\eV}.
\label{eq:Qoneprediction}
\end{equation}
For inverted ordering $Q_\nu<1$ at every finite lightest mass and approaches one only in the degenerate asymptotic limit, so Eq.~\eqref{eq:Qnuone} has no finite IO root.  This ordering selectivity is a genuine consequence of the stated continuation, but the continuation is mechanism dependent: a Majorana operator need not inherit the sterile-Dirac value $Y_R=0$ in this phenomenological feature map.

\subsection{Neutral continuation of \texorpdfstring{$M_H$}{MH}}
For a right-handed neutrino, $Y_R=0$, so a literal continuation of Eq.~\eqref{eq:MH} would set $A_\nu=A_0=0.6966$ and require
\begin{equation}
Q_\nu=\frac{m_1m_3}{m_2^2}=e^{4A_0}=16.22.
\end{equation}
This has no physical solution with the NuFIT central splittings: for normal ordering the kinematically allowed function reaches only $Q_\nu^{\max}=2.930$, while for inverted ordering $Q_\nu<1$.  Thus $M_H$ improves the charged-sector description but fails as a direct neutral-sector extension.  It neither replaces $M_{2\nu}$ nor predicts an absolute neutrino mass without additional neutral-sector structure.

\subsection{Comparison of \texorpdfstring{$M_{2\nu}$, $M_{uL}$, and $M_{dG}$}{M2nu, MuL, and MdG}}
The exact invariant in Eq.~\eqref{eq:R3} fixes the neutrino curvature directly from the central charged-lepton inputs,
\begin{equation}
Q_\ell(M_Z)=\frac{y_ey_\tau}{y_\mu^2}=0.080633.
\label{eq:Qdirect}
\end{equation}
Both $M_{2\nu}$ and $M_{uL}$ use this direct value.  Their neutral prediction is therefore identical, although $M_{uL}$ has the substantially better charged constraint.  A globally fitted M2 curvature would be only a conditional projection and is not used for the central prediction.

\paragraph{Direct identity versus a conditional fitted plane.}
The neutral results summarized after the spectra deliberately use the direct charged-lepton curvature because the model statement being tested is $A_\nu=A_\ell$ (or $A_\nu=A_\ell/5$), not ``$A_\nu$ equals a globally shifted charged fit.''  Nevertheless, the charged feature plane is not exact, so the alternative convention is a useful model-sensitivity check.  Fitting all charged states with $\sth=0.25$ gives
\begin{align}
M_{uL}:&\quad Q_\nu=0.07717,\qquad
\sum m_\nu=0.058908\,\eV\ ({\rm NO}),\quad0.103124\,\eV\ ({\rm IO}),\\
M_{dG}:&\quad Q_\nu=0.61763,\qquad
\sum m_\nu=0.059791\,\eV\ ({\rm NO}),\quad0.16660\,\eV\ ({\rm IO}).
\end{align}
For $M_{uL}$, solving from independent nonsingular charged-sector pairs gives $Q_\nu=0.05190$--$0.08063$, which changes the IO sum from about $0.1017$ to $0.1034\,\eV$ and leaves the NO sum essentially at its oscillation floor.  For $M_{dG}$ the corresponding range $Q_\nu=0.604$--$0.720$ moves the IO sum from about $0.164$ to $0.197\,\eV$.  The many digits in the central spectra displayed immediately below reproduce the algebra from the stated central inputs; they must not be read as comparable physical precision.
For normal ordering, let $x=m_1^2$.  With Eq.~\eqref{eq:nufit}, Eq.~\eqref{eq:R3} becomes
\begin{equation}
x(x+\Delta m_{31}^2)=Q_\ell^2(x+\Delta m_{21}^2)^2.
\label{eq:NOquad}
\end{equation}
For inverted ordering, with $x=m_3^2$ and $D_{32}=|\Delta m_{32}^2|$,
\begin{equation}
x(x+D_{32}-\Delta m_{21}^2)=Q_\ell^2(x+D_{32})^2.
\label{eq:IOquad}
\end{equation}
The positive roots give
\begin{align}
\mathrm{NO}:\quad &(m_1,m_2,m_3)=(1.213\times10^{-4},\,8.682\times10^{-3},\,5.0110\times10^{-2})\,\eV,\\
&\sum m_\nu=0.058914\,\eV,\\[2mm]
\mathrm{IO}:\quad &(m_1,m_2,m_3)=(4.9238\times10^{-2},\,4.9998\times10^{-2},\,4.094\times10^{-3})\,\eV,\\
&\sum m_\nu=0.103329\,\eV.
\label{eq:M2nupred}
\end{align}
For $M_{dG}$, Eq.~\eqref{eq:MdGneutral} gives
\begin{equation}
Q_\nu=Q_\ell^{1/5}=0.604369.
\end{equation}
The corresponding central spectra are
\begin{align}
\mathrm{NO}:\quad &(m_1,m_2,m_3)=(9.191\times10^{-4},\,8.730\times10^{-3},\,5.0118\times10^{-2})\,\eV,
&\sum m_\nu&=0.059767\,\eV,\\
\mathrm{IO}:\quad &(m_1,m_2,m_3)=(6.2295\times10^{-2},\,6.2897\times10^{-2},\,3.8381\times10^{-2})\,\eV,
&\sum m_\nu&=0.163573\,\eV.
\label{eq:MdGpred}
\end{align}

The unnumbered table below summarizes the neutral restrictions and spectra at the point where all entries have been defined.

\begin{table}[H]
\centering
\caption*{Neutral consequences of M2 and the leading projector models.  Central values use the direct charged-lepton curvature rather than fitted charged projections.}
\small
\begin{tabularx}{\textwidth}{@{}ll>{\raggedright\arraybackslash}X>{\raggedright\arraybackslash}X@{}}
\toprule
charged model & neutral restriction & NO: $m_{\rm lightest},\sum m_\nu$ & IO: $m_{\rm lightest},\sum m_\nu$\\
\midrule
M2 $\to M_{2\nu}$ & $A_\nu=A_\ell$ & $1.213\!\times\!10^{-4},\ 0.058914$ & $0.004094,\ 0.103329$\\
$M_{uL}$ & $A_\nu=A_\ell$ & $1.213\!\times\!10^{-4},\ 0.058914$ & $0.004094,\ 0.103329$\\
$M_{B+uL}$ & $A_\nu=A_\ell$, $B_\nu$ free & $1.213\!\times\!10^{-4},\ 0.058914$ & $0.004094,\ 0.103329$\\
$M_{B+dLR}$ & $A_\nu=0$, $B_\nu$ free & $0.001551,\ 0.060504$ & no finite root\\
$M_{B+dLR}^{s14}$ & $A_\nu=0$, $B_\nu$ free & $0.001551,\ 0.060504$ & no finite root\\
$M_{B+\ell T}^{[P_u-3Y_L]}$ & $A_\nu=A_\ell$ & $1.213\!\times\!10^{-4},\ 0.058914$ & $0.004094,\ 0.103329$\\
$M_{B+\ell T}^{[2P_\ell+T_{3L}]}$ & $A_\nu=5A_\ell/3$ & $2.264\!\times\!10^{-5},\ 0.058814$ & $0.000762,\ 0.099671$\\
$M_{B,LR}$ & $A_\nu=0$, $B_\nu$ free & $0.001551,\ 0.060504$ & no finite root\\
$M_{2+B+dLR}$ & $A_\nu=0$, $B_\nu$ free & $0.001551,\ 0.060504$ & no finite root\\
$M_{2+B+dLR}^{[A_\ell/3]}$ & $A_\nu=0$, $B_\nu$ free & $0.001551,\ 0.060504$ & no finite root\\
$M_{2+B+dLR}^{[\delta=1/3]}$ & $A_\nu=0$, $B_\nu$ free & $0.001551,\ 0.060504$ & no finite root\\
$M_{2+B+dLR}^{[B_d/16]}$ & $A_\nu=0$, $B_\nu$ free & $0.001551,\ 0.060504$ & no finite root\\
$M_{dG}$ & $A_\nu=A_\ell/5$ & $0.000919,\ 0.059767$ & $0.03838,\ 0.16357$\\
$M_B$ & $A_\nu$ unrestricted & no prediction & no prediction\\
$M_H$ & $Q_\nu=16.22$ & no physical root & no physical root\\
\bottomrule
\end{tabularx}
\end{table}

\begin{figure}[H]
\centering
\includegraphics[width=0.98\textwidth]{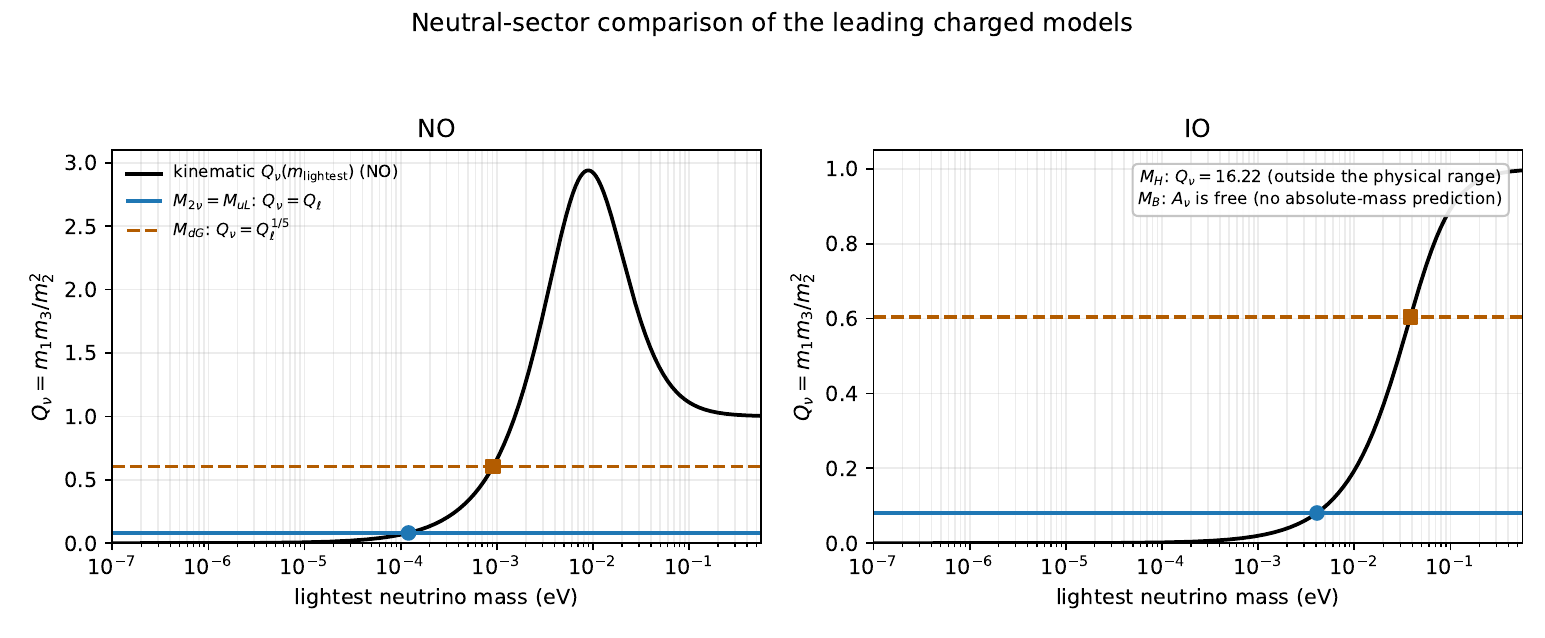}
\caption{Neutral-sector comparison with NuFIT~6.1 central splittings.  Blue intersections correspond to $A_\nu=A_\ell$; the orange dashed line shows $M_{dG}$.  The green dotted line is the $Q_\nu=1$ rule of $M_{B+dLR}$ and $M_{B,LR}$: it intersects only the NO branch at finite mass.  The charged model $M_B$ alone leaves $A_\nu$ free, whereas $M_H$ lies outside the physical range.}
\label{fig:neutralcomparison}
\end{figure}

The narrow propagated input errors would understate the model uncertainty.  The relevant uncertainty is dominated by the uncalibrated structural discrepancy, the selected feature plane, and the neutral matching convention.  The pairwise ranges above provide a transparent diagnostic of this dependence, but they are not confidence intervals.

The predictions should ultimately be tested with full likelihoods from beta decay, cosmology, and, for Majorana neutrinos, neutrinoless double-beta decay \cite{KATRIN2025,DESI2025,PDG2026}.  The DESI DR2 analysis gives $\sum m_\nu<0.0642\,\eV$ (95\%) for one $\Lambda$CDM data combination.  The $Q_\nu=1$ value $0.060504\,\eV$ and the $M_{uL}$ NO value $0.058914\,\eV$ are both compatible but close to that bound, whereas either IO branch is disfavored within this cosmological model.  The same analysis finds $\sum m_\nu<0.163\,\eV$ in a $w_0w_a$CDM extension, which permits the $M_{uL}$ IO value and places the direct $M_{dG}$ IO value near the boundary \cite{DESI2025}.  This strong model dependence is why we report compatibility statements rather than a universal exclusion.

\section{Joint audit of complete charged and neutrino predictions}\label{sec:jointaudit}

\subsection{Why the selection criterion must be strengthened}

A small residual after fitting all nine masses answers ``does the subspace pass close to the data?'', not ``can it recover a hidden mass?''  The final shortlist therefore requires simultaneously: (i) a charged relation evaluated at a fixed, stated renormalization scale; (ii) a unique LOO prediction for all nine states; (iii) a rule for $A_\nu$ without a new neutrino parameter; and (iv) a physical root whose sum does not exceed, or only slightly exceeds, the stated model-dependent cosmological bounds.  This criterion does not turn a post-selected scan into an independent experiment, but it prevents interpolation from being called prediction.

\subsection{The complete \texorpdfstring{$M_{B+\ell T}$}{MB+lT} model}

The selective audit of four simple curvature-ray families identifies
\begin{equation}
\boxed{M_{\ell T}:\quad A_s=a(P_{u,s}-3Y_{L,s})
=a(2P_{\ell,s}+T_{3L,s}),\qquad A_\ell:A_d:A_u=3:-1:1.}
\label{eq:MlT}
\end{equation}
The two forms in Eq.~\eqref{eq:MlT} generate the same three-component vector on the charged sectors.  The model has seven parameters, complete LOO coverage, and supplements the original $A_\ell+3A_d=0$ relation with $A_d+A_u=0$.  Imposing the $M_B$ slope relation at the same time gives
\begin{equation}
\boxed{M_{B+\ell T}:\quad B_s\in\mathrm{span}(|Y_R|,Y_L),\qquad
A_s=a(P_{u,s}-3Y_{L,s}),}
\label{eq:MBLT}
\end{equation}
with three free $I_s$.  This is a six-parameter model with three independent charged relations.  Its maximum conditional shift is $6.81\%$, its worst LOO error is $13.38\%$, and the second-generation errors are $5.44\%$ for $\mu$, $8.08\%$ for $s$, and $10.97\%$ for $c$.  Among the original charge-based feature models, it has the best global LOO stability; the integer-ray extensions below improve it further.

The charged equivalence of the two forms in Eq.~\eqref{eq:MlT} does not continue automatically to neutrinos.  For $P_u-3Y_L$, the neutral row equals the charged-lepton row, so $A_\nu=A_\ell$ and
\begin{equation}
\sum m_\nu=0.058914\,\eV\ ({\rm NO}),\qquad
0.103329\,\eV\ ({\rm IO}).
\end{equation}
For $2P_\ell+T_{3L}$, the neutral value is $5/2$ while the charged-lepton value is $3/2$; hence $A_\nu=5A_\ell/3$ and
\begin{equation}
\sum m_\nu=0.058814\,\eV\ ({\rm NO}),\qquad
0.099671\,\eV\ ({\rm IO}).
\end{equation}
The charged data therefore select one direction but not its ultraviolet operator representation.  We use the explicit $P_u-3Y_L$ form as the primary continuation because it directly extends the logic of $M_{uL}$; the second form is retained as an estimate of operator ambiguity.

\subsection{Six-parameter integer-ray improvements of \texorpdfstring{$M_{B+\ell T}$}{MB+lT}}

The original curvature vector $(3,-1,1)$ is simple and charge motivated, but it is not the only fixed ray that preserves the six-parameter structure.  To test whether its predictive performance can be improved without adding a fitted coefficient, we declare the discrete family
\begin{equation}
\boxed{
M_{B+\ell T_n}:\quad
B_s\in\operatorname{span}(|Y_R|,Y_L),\qquad
A_s=a\bigl[n\mathbf1+2(n+1)P_\ell-(2n+1)P_d\bigr]_s,
}
\label{eq:MBLTn}
\end{equation}
where $n$ is a fixed positive integer, not a continuous fit parameter.  On the three charged sectors this gives
\begin{equation}
A_\ell:A_d:A_u=(3n+2):-(n+1):n.
\label{eq:MBLTnratio}
\end{equation}
The three offsets remain independent, the slope block remains the $M_B$ plane, and the curvature block remains one dimensional.  Hence every fixed-$n$ model has rank six, three charged constraints, common-rescaling invariance, and complete 9/9 LOO identifiability.

The first three rays give
\begin{equation}
n=1:(5,-2,1),\qquad n=2:(8,-3,2),\qquad n=3:(11,-4,3).
\end{equation}
Their numerical performance is substantially better than the original $(3,-1,1)$ ray.  For $n=1$, the maximum conditional shift is $3.92\%$ and the worst LOO error is $8.46\%$.  For $n=2$, they are $1.25\%$ and $7.71\%$; the individual second-generation LOO errors are $3.32\%$ for $\mu$, $2.11\%$ for $s$, and $0.57\%$ for $c$.  For $n=3$, the central-data worst LOO error reaches its family minimum, $4.99\%$, with a $2.25\%$ conditional shift; the corresponding middle-generation errors are $1.89\%$, $3.75\%$, and $3.51\%$.

Two diagnostics prevent us from simply declaring the smallest central number the winner.  First, in a nested LOO test over the declared range $1\le n\le15$, $n$ is reselected using only the eight retained masses.  The other eight select $n=2$ in eight of the nine folds and $n=1$ only when $c$ is held out; the resulting largest prediction error is still $7.71\%$.  Thus the $n=2$ improvement survives the selection step itself.  Second, in 4000 replicas propagated from the quoted marginal input errors, the median worst LOO errors are $14.21\%$ for the original ray, $8.69\%$ for $n=2$, and $7.85\%$ for $n=3$.  The improvement is not a fluctuation of the last displayed digits, although the broad intervals confirm that theory uncertainty dominates.

\begin{figure}[H]
\centering
\includegraphics[width=0.98\textwidth]{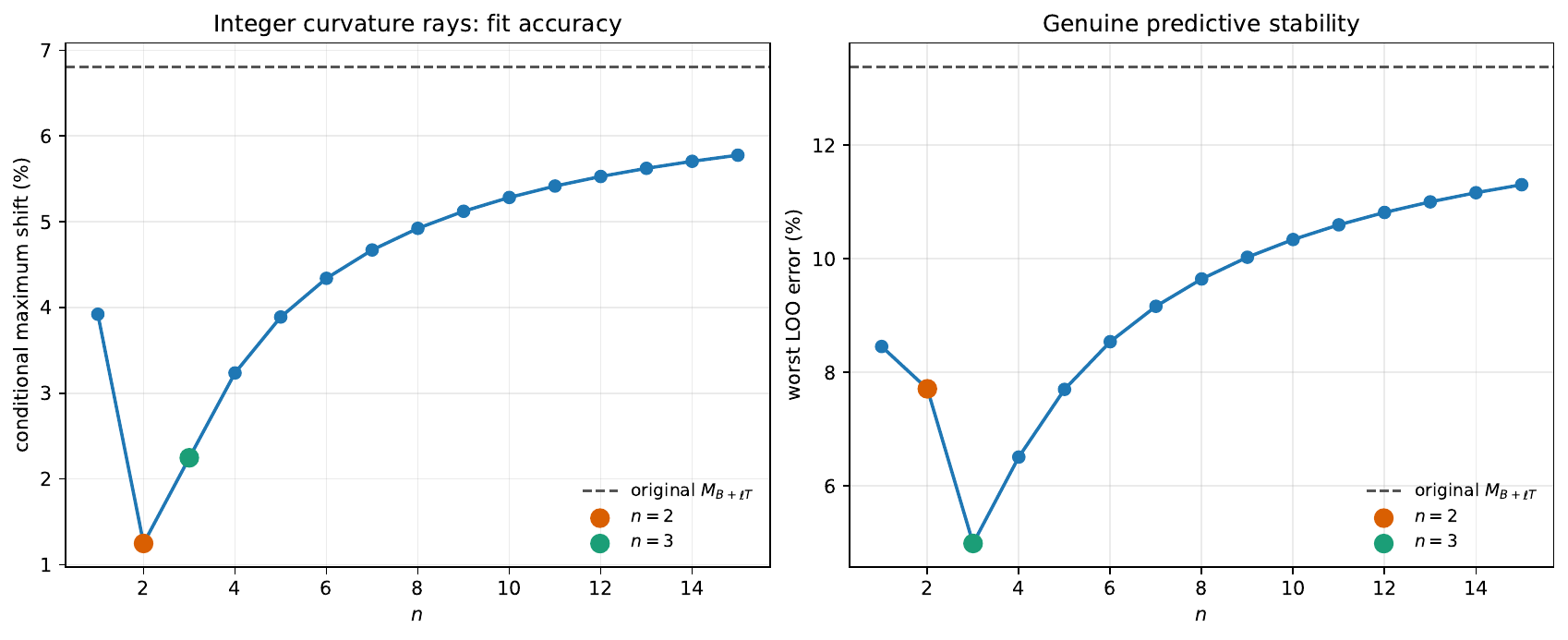}
\caption{Audit of the fixed integer-curvature family in Eq.~\eqref{eq:MBLTn}.  The left panel gives the maximum conditional displacement with all nine masses supplied; the right gives the worst fixed-model LOO error.  Both $n=2$ and $n=3$ improve substantially on the original $M_{B+\ell T}$ ray.  The central-data minimum is $n=3$, whereas nested selection using only the retained masses favors $n=2$.}
\label{fig:MBLTinteger}
\end{figure}

\paragraph{Plain-language reading of Fig.~\ref{fig:MBLTinteger}.}
The dashed line is the original model.  Points below it improve the corresponding diagnostic without adding a parameter.  The sharp dip at $n=2$ on the left means that this ray passes very close to all nine known masses.  The minimum at $n=3$ on the right means that, if this ray is fixed in advance, its worst hidden-mass reconstruction is the smallest for the central inputs.  Because $n=2$ is selected in eight of nine nested folds, we treat it as the conservative predictive extension and $n=3$ as the best central minimax alternative.

For the neutral row used here, $P_\ell=1$ and $P_d=0$, so Eq.~\eqref{eq:MBLTn} gives the same value $3n+2$ as in the charged-lepton sector.  Therefore
\begin{equation}
A_\nu=A_\ell,\qquad
\sum m_\nu=0.058914\,\eV\ ({\rm NO}),\qquad
0.103329\,\eV\ ({\rm IO})
\label{eq:MBLTnnu}
\end{equation}
for every fixed $n$.  The charged improvement therefore does not require an extra neutrino parameter and does not raise the low NO sum.

The family has a natural finite-system reading: a single shell-curvature amplitude is weighted by integer constituent-channel multiplicities that differ between the charged-lepton, down, and up sectors.  A useful decomposition is
\begin{equation}
(3n+2,-n-1,n)=n(3,-1,1)+(2,-1,0),
\label{eq:MBLTtwocorr}
\end{equation}
or, after removing the irrelevant overall normalization, $(3,-1,1)+n^{-1}(2,-1,0)$.  The following construction makes the two-operator statement precise and separates its exact part from the still-conditional ultraviolet dynamics.

\subsubsection{Two operator images and the origin of their signs}\label{sec:twooperators}

For each generation define the three gauge-invariant SM Yukawa monomials
\begin{equation}
 \bm{\mathcal J}_g=
 \begin{pmatrix}
  \overline L_g H e_{Rg}\\[1mm]
  \overline Q_g H d_{Rg}\\[1mm]
  \overline Q_g \widetilde H u_{Rg}
 \end{pmatrix},
 \qquad \widetilde H=i\sigma_2H^\ast ,
\label{eq:YukawaMonomials}
\end{equation}
ordered as $(\ell,d,u)$.  A microscopic interaction that multiplicatively dresses these three monomials has, after confinement and matching, a logarithmic kernel acting on this direct sum.  Two minimal infrared images of such insertions are
\begin{align}
 \widehat{\mathcal Q}_0&=2(P_u-3Y_L),
 & {\cal O}_0&=\sum_g h_g\,\bm q_0^{\trans}\bm{\mathcal J}_g+\text{h.c.},
 \label{eq:O0def}\\
 \widehat{\mathcal Q}_1&=-3(B-L)-P_\ell+P_u,
 & {\cal O}_1&=\sum_g h_g\,\bm q_1^{\trans}\bm{\mathcal J}_g+\text{h.c.},
 \label{eq:O1def}
\end{align}
where $\bm q_i$ is the vector of eigenvalues of $\widehat{\mathcal Q}_i$ on the charged sectors.  The projectors in Eqs.~\eqref{eq:O0def}--\eqref{eq:O1def} are sector labels fixed by SM representations; they are not fitted functions of the masses.  Substitution of the SM quantum numbers gives
\begin{table}[H]
\centering
\caption{Charge derivation of the two curvature operators.  The signs and zeros follow before any Yukawa eigenvalue is inserted.}
\label{tab:operatorcharges}
\small
\begin{tabular}{@{}lrrrrrr@{}}
\toprule
sector $s$ & $Y_{L,s}$ & $(B-L)_s$ & $P_{\ell,s}$ & $P_{u,s}$ & $q_{0s}=2(P_u-3Y_L)$ & $q_{1s}=-3(B-L)-P_\ell+P_u$\\
\midrule
$\ell$ & $-1/2$ & $-1$  & $1$ & $0$ & $+3$ & $+2$\\
$d$    & $+1/6$ & $+1/3$& $0$ & $0$ & $-1$ & $-1$\\
$u$    & $+1/6$ & $+1/3$& $0$ & $1$ & $+1$ & $0$\\
\bottomrule
\end{tabular}
\end{table}
Thus
\begin{equation}
 \boxed{\bm q_0=(3,-1,1),\qquad \bm q_1=(2,-1,0).}
\label{eq:q0q1}
\end{equation}
This answers the sign question without referring to the measured spectrum.  The negative down-sector signs originate in the positive quark value of $Y_L$ for $\widehat{\mathcal Q}_0$ and the positive quark value of $B-L$ for $\widehat{\mathcal Q}_1$; the up-sector zero of $\bm q_1$ is the exact cancellation $-3(1/3)+P_u=0$.  What is not derived by this algebra is why a particular ultraviolet Hamiltonian should generate precisely the combination in Eq.~\eqref{eq:O1def}.  The cancellation is therefore a sharp microscopic selection rule to be tested, not a consequence of the mass fit.

The use of logarithmic coefficients also needs to be explicit.  Let $Y^{(0)}_{s,g}$ denote the smooth matched Yukawa coefficient before the two curvature insertions.  Repeated dressing, or equivalently exponentiation of the connected matching kernel, gives
\begin{equation}
 Y_{s,g}(M_{\rm match})=Y^{(0)}_{s,g}
 \exp\!\left[h_g\left(C_0q_{0s}+C_1q_{1s}\right)\right].
\label{eq:logmatchingkernel}
\end{equation}
To first order this is the usual additive operator matching; the exponential form resums repeated multiplicative insertions and is the form appropriate to $z_{s,g}=\ln Y_{s,g}$.  Equation~\eqref{eq:logmatchingkernel} produces
\begin{equation}
 \bm A=C_0\bm q_0+C_1\bm q_1.
\label{eq:AfromC}
\end{equation}
Choosing $C_0=na$ and $C_1=a$ then yields Eq.~\eqref{eq:MBLTnratio} and the measurable ratio
\begin{equation}
 \boxed{\frac{C_1}{C_0}=\frac1n.}
\label{eq:Cratio}
\end{equation}

\subsubsection{How \texorpdfstring{$n=2$ or $n=3$}{n=2 or n=3} can be fixed before the spectrum}\label{sec:ntopology}

A group or diagrammatic calculation must determine Eq.~\eqref{eq:Cratio}; it cannot merely rename a fitted integer.  A minimal topology-counting completion is obtained as follows.  Suppose the two matched coefficients have the form
\begin{equation}
 C_i=N^{-k_i}\sum_{\alpha=1}^{d_i}\kappa_i^{(\alpha)},
\label{eq:topologycoeff}
\end{equation}
where $k_i$ counts preon-line interchanges, $d_i$ counts symmetry-related contractions or degenerate intermediate states, and $\kappa_i^{(\alpha)}$ is a reduced matrix element.  If an exact ultraviolet symmetry makes all relevant reduced matrix elements equal to one common $\kappa$ and the two insertions occur at the same interchange order, $k_0=k_1$, then
\begin{equation}
 \frac{C_1}{C_0}=\frac{d_1}{d_0}.
\label{eq:degeneracyratio}
\end{equation}
The discrete choice is now made by the state content and diagram topology, before any charged mass is examined:
\begin{table}[H]
\centering
\caption{Two conditional ultraviolet matching classes and the status of the published $SU(15)_p$ construction.  The first two rows are hypotheses with explicit pre-spectral selection rules, not claims about Ref.~\cite{AssiEtAl2026}.}
\label{tab:UVclasses}
\footnotesize
\begin{tabularx}{\textwidth}{@{}lccc>{\raggedright\arraybackslash}X@{}}
\toprule
class & $(d_0,d_1)$ & $C_1/C_0$ & ray & ultraviolet requirement fixed before masses\\
\midrule
$\mathcal D_2$ & $(2,1)$ & $1/2$ & $(8,-3,2)$ & two exchange-related orientations of ${\cal O}_0$, one allowed ${\cal O}_1$ route, equal reduced amplitudes enforced by a discrete interchange symmetry\\
$\mathcal D_3$ & $(3,1)$ & $1/3$ & $(11,-4,3)$ & a cyclic symmetry of three constituent slots gives three equal ${\cal O}_0$ contractions, while a boundary or spurion selection leaves one ${\cal O}_1$ contraction\\
published $SU(15)_p$ flavor theory & not fixed & not fixed & not fixed & three-preon prebaryons and integer preon-line interchange counting exist, but inequivalent spurions and independent nonperturbative coefficients prevent Eqs.~\eqref{eq:degeneracyratio} and \eqref{eq:Cratio} from following automatically \cite{AssiEtAl2026}\\
\bottomrule
\end{tabularx}
\end{table}
In class $\mathcal D_2$, $n=2$ is a group-theoretic degeneracy statement; in class $\mathcal D_3$, $n=3$ is a cyclic-topology statement.  The masses are used only afterwards to test the resulting rays.  The first class is consistent with the nested-LOO preference for $n=2$, while the second is consistent with the fixed-central-data minimax value $n=3$.  Neither agreement retroactively proves the assumed symmetry.  A published $SU(15)_p$ model contains three-preon states and several four- and six-prebaryon matching topologies, but its reduced coefficients associated with inequivalent spurions are free and need not be equal \cite{AssiEtAl2026}.  Consequently the existing theory supplies the vocabulary for Eq.~\eqref{eq:topologycoeff}, not the required equality of its matrix elements.

If $k_1=k_0+1$ instead, Eq.~\eqref{eq:topologycoeff} would give $C_1/C_0=(d_1/d_0)/N$, not $1/n$.  This observation rules out a common but incorrect shortcut: one cannot explain $n=2$ or $3$ merely by saying that ${\cal O}_1$ has one additional preon-line exchange in an $SU(15)$ theory.  The observed rays require a same-order degeneracy ratio, or another independently calculated group factor.  This is a useful falsifiability condition on any microscopic completion.

\subsection{A targeted \texorpdfstring{$M_{B+dLR}$}{MB+dLR} improvement that predicts \texorpdfstring{$s$}{s}}

The non-identifiability of $s$ can be removed without adding a parameter by tying the full down-sector curvature to its slope.  A basis-independent form of the targeted model is
\begin{equation}
\boxed{M_{B+dLR}^{s14}:\quad
A_s=\lambda\!\left(Y_{L,s}Y_{R,s}+\frac1{18}P_{d,s}\right)
+\frac{B_d}{14}P_{d,s},\qquad B_s\in\mathrm{span}(|Y_R|,Y_L).}
\label{eq:MBdLRs14}
\end{equation}
The first direction is $(1/2,0,1/9)$, so it preserves $2A_\ell-9A_u=0$ without supplying an independent down curvature.  The second term imposes $A_d=B_d/14$.  Since $B_d$ is already determined by the two coefficients of $M_B$, the parameter count falls from seven to six and $s$ becomes identified.

For the down triplet itself, the new rule can be read without matrix algebra:
\begin{equation}
A_d=\frac{B_d}{14}
\quad\Longleftrightarrow\quad
4\ln y_d-7\ln y_s+3\ln y_b=0
\quad\Longleftrightarrow\quad
\boxed{y_s^{\rm pred}=(y_d^4y_b^3)^{1/7}.}
\label{eq:spred}
\end{equation}
Thus the strange-quark mass becomes a weighted geometric mean of the $d$ and $b$ masses.  Their central values give $y_s^{\rm pred}=3.0465\times10^{-4}$ instead of $3.0600\times10^{-4}$, a direct difference of $0.44\%$.  The full simultaneous LOO test shifts the endpoints slightly through the $M_B$ relation and gives the $0.51\%$ value quoted below.

At $M_Z$ this model gives $\chi^2_{25\%}=0.01370$, $d_{\rm subspace}=0.02619$, a maximum conditional shift of $1.99\%$, complete 9/9 coverage, and an $s$-specific LOO error of only $0.51\%$.  The worst LOO error over the full set is $25.06\%$ and occurs for $e$, so targeted accuracy for $s$ does not make the model globally most stable.  The neighboring simple choice $A_d=B_d/15$ gives $2.78\%$ for $s$ and lowers the overall maximum to $23.89\%$.

The central inputs give $A_d/B_d=0.070795$, compared with $1/14=0.071429$.  Propagating only the quoted marginal errors gives $0.070804\pm0.002173$; among $1/n$ candidates with $8\le n\le30$, $1/14$ is nearest in $73.2\%$ of draws.  This shows numerical input stability but does not remove post-selection: the number 14 was noticed using the same $s$ mass that the model subsequently ``predicts.''  The relation is also scale dependent.  The next subsection therefore asks the more restrictive question: can a theoretically motivated $1/N=1/15$ matching factor evolve into the observed low-energy number without replacing $N=15$ by an unexplained $N-1=14$?

\subsubsection{Matching and RG evolution: how \texorpdfstring{$1/15$}{1/15} can appear as \texorpdfstring{$1/14$}{1/14} at \texorpdfstring{$M_Z$}{MZ}}\label{sec:RG1514}

Define $r_d(\mu)=A_d(\mu)/B_d(\mu)$ from the running down-type Yukawa eigenvalues.  With $t=\ln\mu$ and $\gamma_i=\dd\ln y_i/\dd t$, Eq.~\eqref{eq:inverseIBA} gives
\begin{align}
 \frac{\dd A_d}{\dd t}&=\frac14\left(\gamma_d-2\gamma_s+\gamma_b\right),
 &
 \frac{\dd B_d}{\dd t}&=\frac12\left(\gamma_b-\gamma_d\right),
 \label{eq:RGAB}\\
 \frac{\dd r_d}{\dd t}&=
 \frac{B_d\,\dd A_d/\dd t-A_d\,\dd B_d/\dd t}{B_d^2}.
 \label{eq:RGratio}
\end{align}
The unequal anomalous dimensions of $d,s,b$ therefore renormalize the curvature-to-slope ratio even if a UV matching condition fixes it to a simple group factor.  This is the specific mechanism that can move a rational coefficient; no change in the number of preon colors is required.

Using the two-loop SM running eigenvalues tabulated in Ref.~\cite{Antusch2026}, we obtain
\begin{table}[H]
\centering
\caption{Reverse matching audit of the down-sector ratio and the two-operator curvature direction.  The last column is an unconstrained projection diagnostic, not the nested-LOO model ranking.}
\label{tab:RG1514}
\small
\begin{tabular}{@{}lrrl@{}}
\toprule
common scale $\mu$ & $A_d/B_d$ & $n_{\rm eff}=C_0/C_1$ & nearest relevant reference\\
\midrule
$M_Z$              & $0.070795$ & $1.730$ & $0.887\%$ below $1/14$\\
$10^3\,\GeV$       & $0.069710$ & $1.648$ & between $1/14$ and $1/15$\\
$3\times10^3\,\GeV$& $0.069424$ & $1.595$ & between $1/14$ and $1/15$\\
$10^5\,\GeV$       & $0.068177$ & $1.569$ & between $1/14$ and $1/15$\\
$10^7\,\GeV$       & $0.066486$ & $1.484$ & $0.271\%$ below $1/15$\\
$10^9\,\GeV$       & $0.066091$ & $1.454$ & close to $1/15$\\
$10^{16}\,\GeV$    & $0.063007$ & $1.323$ & close to $1/16$\\
\bottomrule
\end{tabular}
\end{table}
Here $n_{\rm eff}$ is defined by projecting the direct curvature vector $\bm A=(A_\ell,A_d,A_u)^{\trans}$ onto the operator basis of Eq.~\eqref{eq:q0q1}:
\begin{equation}
 \begin{pmatrix}C_0\\ C_1\end{pmatrix}_{\!\rm LS}
 =({\mathsf Q}^{\trans}{\mathsf Q})^{-1}{\mathsf Q}^{\trans}\bm A,
 \qquad {\mathsf Q}=(\bm q_0\ \bm q_1),
 \qquad n_{\rm eff}=\frac{C_0}{C_1}.
\label{eq:neff}
\end{equation}
It records the continuous direction selected by the running spectrum.  It must not be confused with a microscopic integer: class $\mathcal D_2$ or $\mathcal D_3$ fixes $n$ first, and the corresponding RG-evolved ray is then tested.  The low-scale value $n_{\rm eff}=1.73$ lies nearer the $n=2$ topology than the $n=3$ topology, consistently with the nested-LOO preference, but the drift in Table~\ref{tab:RG1514} shows that a fixed ray is not an SM RG invariant.

Log-linear interpolation of the two-loop entries around the crossing gives
\begin{equation}
 r_d(\Lambda_{15})=\frac1{15},\qquad
 \log_{10}\!\frac{\Lambda_{15}}{\GeV}\simeq6.786,\qquad
 \boxed{\Lambda_{15}\simeq6.1\times10^6\,\GeV.}
\label{eq:Lambda15}
\end{equation}
Between $10^7\,\GeV$ and $M_Z$ the tabulated trajectory multiplies the ratio by
\begin{equation}
 \frac{r_d(M_Z)}{r_d(10^7\,\GeV)}=1.06481.
\label{eq:RGfactor1514}
\end{equation}
Acting with this factor on an exact UV value $1/15$ gives $0.070987$, only $0.62\%$ below $1/14$.  Equations~\eqref{eq:RGAB}--\eqref{eq:RGfactor1514} therefore supply a concrete answer to the denominator question: the microscopic suppression may remain $1/N=1/15$, while non-universal SM running of the three down-type Yukawas makes its infrared image numerically close to $1/14$.  There is no need to postulate that the number of preon colors has dynamically changed from 15 to 14.

This is nevertheless a conditional matching result, not a prediction of the matching scale.  The scale in Eq.~\eqref{eq:Lambda15} was located from the running observed spectrum.  A genuine derivation must predict $\Lambda_{\rm match}$ near $10^{6}$--$10^{7}\,\GeV$, compute threshold effects, and reproduce the boundary condition before $m_s$ is used.  If it predicts a substantially different scale, or threshold corrections with the wrong sign, the $1/15\to1/14$ interpretation fails.  A literal $1/14$ may alternatively arise from a rank-fourteen active projector $P_{\rm act}=\mathbf1_{15}-|f\rangle\langle f|$, normalized as $P_{\rm act}/\mathrm{tr}\,P_{\rm act}$, but that option requires an independently derived forbidden preon channel.  No such channel has yet been identified in the published $SU(15)_p$ flavor construction \cite{AssiEtAl2026}; the RG explanation is therefore the more economical conditional mechanism.

Its neutral continuation is unchanged: $P_d=0$ and $Y_R=0$ on the adopted neutral row, so $A_\nu=0$, $Q_\nu=1$, and $\sum m_\nu=0.060504\,\eV$ for NO, with no finite IO root.

\begin{figure}[H]
\centering
\includegraphics[width=0.94\textwidth]{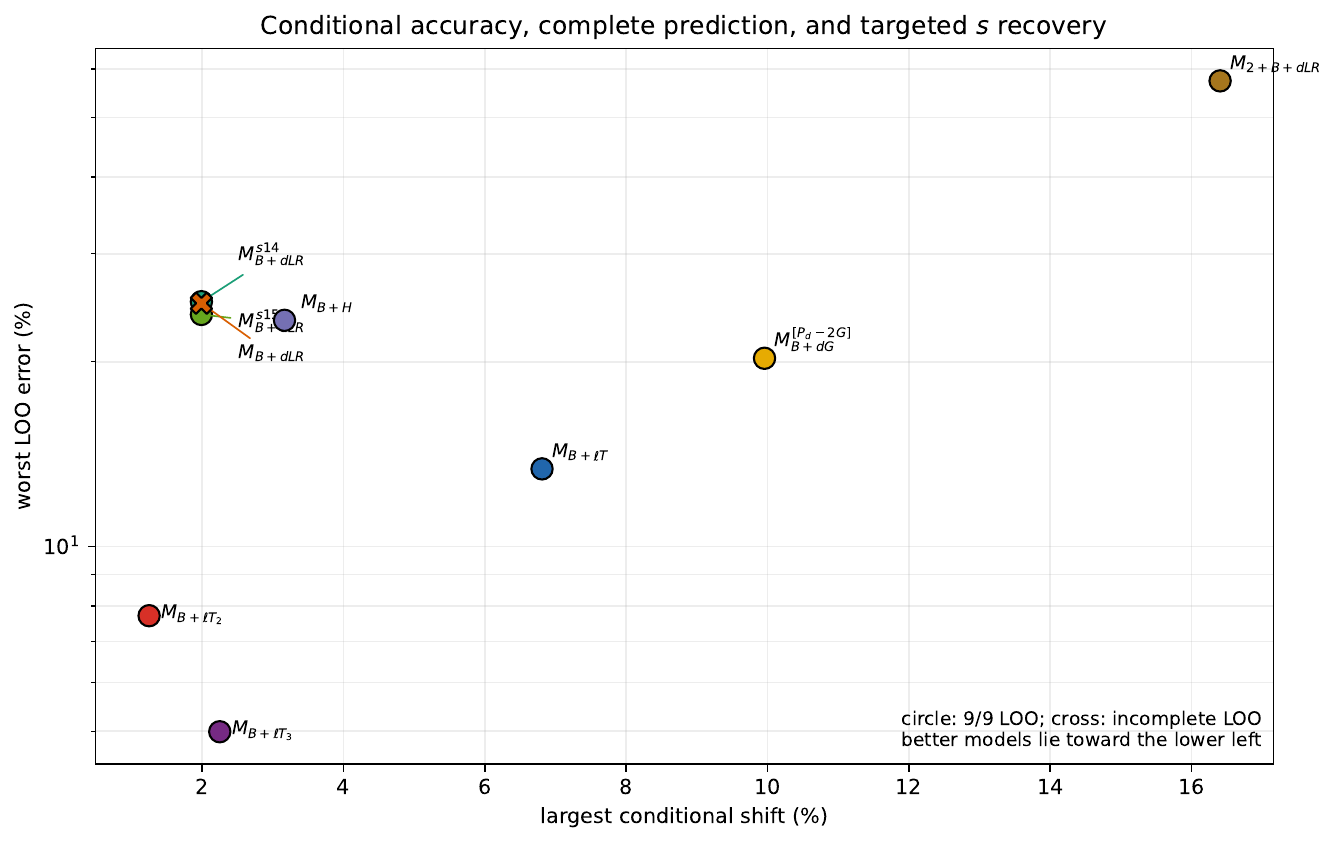}
\caption{Joint comparison of conditional accuracy and LOO prediction.  The $M_{B+\ell T_2}$ and $M_{B+\ell T_3}$ points lie well below and to the left of the original $M_{B+\ell T}$: both improve the complete nine-mass prediction without adding a parameter.  $M_{B+dLR}^{s14}$ remains the most accurate targeted reconstruction of $s$, but not the most stable global model.}
\label{fig:jointtradeoff}
\end{figure}

\paragraph{Plain-language reading of Fig.~\ref{fig:jointtradeoff}.}
$M_{B+dLR}$ passes close to the known points, but the cross shows that it fails to predict one mass.  The rule $A_d=B_d/14$ changes the cross into a circle: all nine masses are then predicted, and $s$ itself is recovered to $0.51\%$.  The point remains high because the worst state becomes the electron.  The simple $M_{B+\ell T}$ is lower, so its errors are more even.  Its two integer improvements now appear directly on the same plot: $M_{B+\ell T_2}$ moves strongly toward the lower left, while $M_{B+\ell T_3}$ gives the lowest central-data worst error.  The plot alone does not include the cost of choosing $n$; the nested selection in Fig.~\ref{fig:MBLTinteger} is why $n=2$, rather than the visually lowest $n=3$, is retained as the conservative favorite.

For completeness, the final table retains one gauge-quadratic control that was not competitive in the global ranking:
\begin{equation}
M_{B+dG}^{[P_d-2G]}:\qquad
B_s\in\operatorname{span}(|Y_R|,Y_L),\qquad
A_s=a(P_{d,s}-2G_s),\qquad G_s=Y_{L,s}^2+Y_{R,s}^2.
\label{eq:MBdGray}
\end{equation}
It is a six-parameter, complete-coverage model.  Its neutral row gives $A_\nu=A_\ell/5$, as in $M_{dG}$, so it provides a useful control with a substantially larger IO sum rather than a leading candidate.

\begin{table}[H]
\centering
\caption{Updated shortlist after the joint and targeted follow-up audits.  Sums are shown as NO; IO in eV.  ``None'' means that no finite physical root exists.}
\label{tab:jointshortlist}
\footnotesize
\begin{tabularx}{\textwidth}{@{}lrrrrc>{\raggedright\arraybackslash}X@{}}
\toprule
model & par. & shift & worst LOO & LOO($s$) & coverage & $\sum m_\nu$ and final role\\
\midrule
$M_{B+\ell T_2}$ & 6 & 1.25\% & 7.71\% & 2.11\% & 9/9 & $0.058914;\ 0.103329$; conservative favorite, selected in 8/9 nested folds; normalized correction $\tfrac12(2,-1,0)$\\
$M_{B+\ell T_3}$ & 6 & 2.25\% & 4.99\% & 3.75\% & 9/9 & $0.058914;\ 0.103329$; best central minimax ray, but more selection sensitive; correction $\tfrac13(2,-1,0)$\\
$M_{2+B+dLR}^{[A_\ell/3]}$ & 6 & 1.86\% & 3.82\% & 2.02\% & 9/9 & $0.060504;$ none; strongest operator-matching clue within the $dLR$ family\\
$M_{2+B+dLR}^{[\delta=1/3]}$ & 6 & 1.91\% & 5.78\% & 2.72\% & 9/9 & $0.060504;$ none; best shell-center interpretation, but more exploratory\\
$M_{2+B+dLR}^{[B_d/16]}$ & 6 & 1.95\% & 3.91\% & 2.01\% & 9/9 & $0.060504;$ none; compact quantum-defect parametrization; microscopic origin still open\\
$M_{B+\ell T}^{[P_u-3Y_L]}$ & 6 & 6.81\% & 13.38\% & 8.08\% & 9/9 & $0.058914;\ 0.103329$; simplest charge-based complete model\\
$M_{B+dLR}^{s14}$ & 6 & 1.99\% & 25.06\% & 0.51\% & 9/9 & $0.060504;$ none; best targeted $s$ recovery, but post-selected\\
$M_{B+dLR}$ & 7 & 1.99\% & 24.92\% & free & 8/9 & $0.060504;$ none; most accurate conditional NO model, not a complete predictor\\
$M_{B+dG}^{[P_d-2G]}$ & 6 & 9.96\% & 20.27\% & 13.50\% & 9/9 & $0.059767;\ 0.163573$; complete control; the dedicated audit did not improve it materially\\
$H[1\!:\!-2]$ & 6 & 6.81\% & 13.38\% & 8.08\% & 9/9 & literal $H$: none; operator-equivalent continuations exist, but the dedicated audit found no robust joint charged-plus-neutral improvement\\
\bottomrule
\end{tabularx}
\end{table}

\paragraph{Plain-language reading of Table~\ref{tab:jointshortlist}.}
If one fixed central-data model is required to minimize the largest charged-mass error, $M_{2+B+dLR}^{[A_\ell/3]}$ and $M_{2+B+dLR}^{[B_d/16]}$ now give the smallest worst LOO errors within the $dLR$ branch, while $M_{B+\ell T_3}$ remains the strongest minimax point of the older integer family.  If the discrete choice itself is included in the validation, $M_{B+\ell T_2}$ is still the safer conservative conclusion because eight of nine nested folds select it without seeing the hidden mass.  The original $M_{B+\ell T}$ remains the simplest charge-based complete model.  If the priority is exceptionally accurate recovery of $s$ alone, $M_{B+dLR}^{s14}$ remains the strongest targeted hypothesis, but its coefficient must still be derived independently.  The seven-parameter $M_{B+dLR}$ must not be described as a predictor of all nine masses.

\subsubsection{Structurally motivated repairs of \texorpdfstring{$M_{2+B+dLR}$}{M2+B+dLR}}

The poor performance of $M_{2+B+dLR}$ is not caused by the $dLR$ curvature plane itself, but by the rigid offset tie $I_\ell=I_d$.  A natural next step is therefore to retain the successful $M_B$ slope plane and the same $dLR$ curvature plane, while replacing that offset rule by a more structured six-parameter relation.  Three such repairs were tested in detail.  The first is
\begin{equation}
 I_\ell-I_d=\frac{1}{3}A_\ell .
 \label{eq:m2star_offset_curvature}
\end{equation}
At first sight the factor $1/3$ looks like another fitted rational number.  Within the present quantum-number dictionary, however, it can be absorbed into one cross-block operator.  On the charged sectors $(\ell,d,u)$ define
\begin{align}
 Q_A &\equiv 2Y_LY_R+P_dY_R^2=(1,0,2/9),\\
 Q_B &\equiv P_d(B-L)=(0,1/3,0).
\end{align}
Then a single shared coefficient $\lambda$ multiplying
\begin{equation}
 \mathcal O_\lambda
 =h_g\left(2Y_LY_R+P_dY_R^2\right)-P_d(B-L)
 \label{eq:m2star_crossblock_operator}
\end{equation}
automatically gives
\begin{equation}
 A_\ell=\lambda,\qquad
 A_u=\frac{2}{9}\lambda,\qquad
 \delta I_d=-\frac{1}{3}\lambda,
\end{equation}
and therefore
\begin{equation}
 2A_\ell-9A_u=0,
 \qquad
 I_\ell-I_d-\frac13 A_\ell=0.
\end{equation}
The same six-parameter model can be written in logarithmic space as
\begin{align}
 \ln y_{s,g}={}&
 I_0(P_\ell+P_d)_s+I_u(P_u)_s
 +(g-2)\left[b_R|Y_R|_s+b_L(Y_L)_s\right]
 +A_d h_g(P_d)_s
 \nonumber\\
 &+\lambda\left\{
 h_g\left[2Y_LY_R+P_dY_R^2\right]_s
 -\left[P_d(B-L)\right]_s
 \right\}.
 \label{eq:m2star_effective_operator}
\end{align}
It has rank six and is exactly equivalent to Eq.~\eqref{eq:m2star_offset_curvature} together with $M_B$ and the $dLR$ plane.  Its third parameter-free charged invariant is
\begin{equation}
 R_2^3=Q_\ell,
 \qquad
 R_2\equiv\frac{y_e y_\mu^2 y_\tau}{y_d y_s^2y_b},
 \qquad
 Q_\ell\equiv\frac{y_e y_\tau}{y_\mu^2},
 \label{eq:m2star_invariant}
\end{equation}
or equivalently
\begin{equation}
 \frac{y_e^2y_\mu^8y_\tau^2}{y_d^3y_s^6y_b^3}=1.
\end{equation}
With the central $M_Z$ inputs this fixed model gives $\chi^2=0.0192$, a maximum conditional shift of $1.86\%$, and a worst LOO error of $3.82\%$ with complete 9/9 coverage.

Two additional six-parameter repairs are also useful.  The shifted-center rule
\begin{equation}
 I_\ell-I_d+\frac13(B_\ell-B_d)=0
 \label{eq:m2star_shiftedcenter}
\end{equation}
is equivalent to requiring the smooth lepton and down trends to meet not at $g=2$, but at $g_*=7/3$.  This admits a literal shell reading in a two-dimensional harmonic-oscillator toy spectrum, where the first three shell degeneracies are $1:2:3$ and the degeneracy-weighted center is $\bar g=7/3$.  Numerically it gives $\chi^2=0.0286$, a $1.91\%$ maximum conditional shift, and a $5.78\%$ worst LOO error.

The third repair,
\begin{equation}
 I_\ell-I_d+\frac{1}{16}B_d=0,
 \label{eq:m2star_bd16}
\end{equation}
may be read as a small down-sector shell-coordinate defect, $x_d=(g-2)+1/16$.  It gives $\chi^2=0.0155$, a $1.95\%$ maximum conditional shift, and a $3.91\%$ worst LOO error.  A nearby theoretically motivated neighbor, $B_d/15$, remains viable but slightly weaker, with $\chi^2=0.0142$, a $2.02\%$ maximum conditional shift, and a $4.73\%$ worst LOO error.

All three repairs share the same neutral continuation as $M_{B+dLR}$ and $M_{2+B+dLR}$: on the adopted neutral row, $P_d=0$ and $Y_R=0$, so $A_\nu=0$, $Q_\nu=1$, and $\sum m_\nu=0.060504\,\eV$ for NO only.  Their physical status remains exploratory.  The operators were identified after examining the same charged masses, their relations are strongly scale sensitive under SM running, and the shared coefficient $\lambda$ in Eq.~\eqref{eq:m2star_crossblock_operator} has not yet been derived from a microscopic preon Hamiltonian or matching calculation.

\begin{figure}[H]
\centering
\includegraphics[width=0.98\textwidth]{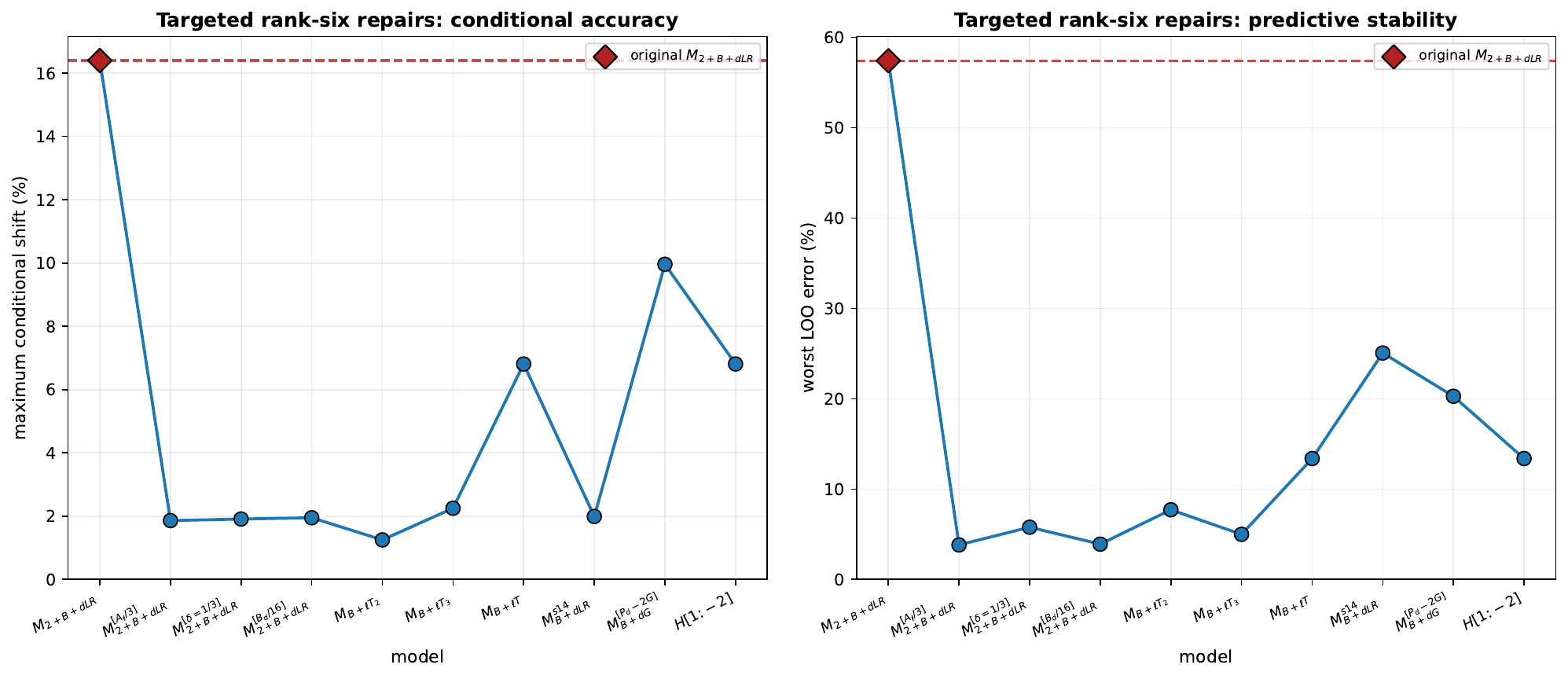}
\caption{Updated charged-model comparison in the visual style of Fig.~\ref{fig:MBLTinteger}.  Left: maximum conditional shift.  Right: worst fixed-model LOO error.  The original $M_{2+B+dLR}$ result is now shown explicitly as the dark-red diamond at $(16.40\%,57.40\%)$ as well as by the dashed reference lines.  The three repaired $M_{2+B+dLR}$ models lie dramatically below and to the left of that point.  Figure~\ref{fig:MBLTinteger} is retained to show the earlier integer-ray improvement of $M_{B+\ell T}$.}
\label{fig:jointtradeoffrepairs}
\end{figure}

\paragraph{Plain-language reading of Fig.~\ref{fig:jointtradeoffrepairs}.}
The dark-red diamond at the upper right is the previously omitted original $M_{2+B+dLR}$ point.  Its coordinates mean that enforcing the uncorrected model requires a $16.40\%$ largest conditional displacement and produces a $57.40\%$ worst LOO error.  The three repaired descendants sit near the lower left, showing directly that the poor performance of the original model came mainly from the offset rule $I_\ell=I_d$, not from the entire $M_2+B+dLR$ construction.  Among them, the $A_\ell/3$ and $B_d/16$ repairs give the lowest worst LOO errors, while the shifted-center rule is slightly weaker numerically but has the cleanest shell-center interpretation.  The older $M_{B+\ell T_2}$ and $M_{B+\ell T_3}$ points remain competitive reference models, whereas $M_{B+dG}^{[P_d-2G]}$ and $H[1\!:\!-2]$ still lie visibly higher in at least one panel.

\paragraph{Brief outcome of the auxiliary $M_{B+dG}$ and $M_{B+H}$ audits.}
The dedicated follow-up audits do not materially improve either auxiliary family as a complete charged-plus-neutral model.  For $M_{B+dG}$, the simple global ray $P_d-2G$ remains the most useful six-parameter control: a high-complexity post-selected ray $33P_d-64G$ lowers the worst central-data LOO error only from $20.27\%$ to $20.14\%$, while the selective ray $13P_d-20G$ recovers $s$ to $0.048\%$ but raises the worst LOO error to $34.67\%$; the cross-rule $A_d=B_d/14$ gives $0.51\%$ for $s$ but $54.72\%$ for $\mu$.  For $M_{B+H}$, charged-only deformations can look numerically impressive---for example $|Y_R|-T_{3L}/40$ gives a $0.527\%$ maximum shift and a $4.10\%$ fixed LOO error---but their literal neutral continuation still yields no physical neutrino root.  The best simple complete $H$ ray remains $1-2|Y_R|$, with $6.81\%$ shift and $13.38\%$ worst LOO, and the low neutrino sums arise only after replacing the literal $H$ continuation by a different neutral operator.

\section{Relation to previous work and scope of novelty}\label{sec:prior}
Empirical relations among fermion masses have a long history.  Koide-type formulae compare special symmetric functions of a triplet and have been studied for running charged-lepton, quark, and neutrino masses \cite{Koide1983,XingZhang2006}.  Kr\'olikowski proposed a unified empirical formula for quark and lepton spectra with compositeness motivation \cite{Krolikowski2013}; Gao and Li compared related empirical constructions \cite{GaoLi2016}.  Other studies connect hierarchy measures to flavor mixing \cite{HollikSaldana2015,YangZhangFeng2024}, while neutrino sum rules combine a model relation with two oscillation splittings to predict an absolute scale \cite{CentellesChulia2024,HuangEtAl2017}.  A recent $\mathcal A_4$ type-II seesaw construction illustrates the stronger standard required of a microscopic sum rule: the mass relation is derived together with the ordering, mixing correlations, Majorana phases, and lepton-number-violating observables \cite{CentellesKumar2026}.  Our relation is intentionally more limited because it is presently an eigenvalue-level hypothesis without such an operator completion.

The present work differs in its design-matrix audit and exhaustive projector comparison.  For each fixed ansatz the predictive content is identified from the left-null space before interpreting the numerical fit, and exact triplet interpolation is excluded from the count of predictions.  The nested audit goes beyond selecting one rank-eight hyperplane: it asks which second-generation masses become identifiable at ranks seven and six and checks the distinction by LOO reconstruction.  The resulting $M_{B+dLR}$ and $M_{B,LR}$ rules add an ordering-selective $Q_\nu=1$ continuation to the previously identified $A_\nu=A_\ell$ and $A_\nu=A_\ell/5$ possibilities.

A particularly close recent terminology appears in a non-peer-reviewed 2026 preprint that assigns charged fermions to a discrete weighted lattice and obtains an exponential hierarchy \cite{ZhangHuZhang2026}.  Its construction does not use the projector-plane rank audit or the invariants derived here.  The comparison is included to make the novelty claim narrower and more transparent.

\subsection*{Connection with current preon dynamics}

Several recent developments make the comparison more concrete than a generic appeal to compositeness.  The chiral $SU(15)_p$ construction of Assi and Dobrescu embeds SM quarks and leptons as composite prebaryons and obtains three generations from the preon-flavor structure \cite{Dobrescu2022,AssiDobrescu2025}.  Its 2026 flavor completion uses two $SU(4)_F$-breaking spurions and a benchmark with order-one nonperturbative coefficients to reproduce all six quark masses, three charged-lepton masses, and CKM mixing \cite{AssiEtAl2026}.  In that construction the quark and lepton Yukawa matrices share flavor spurions, while a right-handed isospin symmetry correlates the up and down matrices \cite{AssiEtAl2026}.  This is qualitatively aligned with our finding that a common slope plane and a single curvature amplitude can constrain all three charged sectors.  It is not an equality of models: the $SU(15)_p$ spurions are matrices with mixing information, whereas our rays constrain eigenvalues only.

The same $SU(15)_p$ analysis offers a second, more specific comparison.  Its low-energy amplitudes are organized as $(1/N)^k$, where $k$ counts preon-line interchanges and each interchange costs approximately $1/N=1/15$ \cite{AssiEtAl2026}.  It also contains three-preon prebaryons and several four- and six-prebaryon Yukawa topologies.  These facts motivate the bookkeeping in Eq.~\eqref{eq:topologycoeff}, but they do not fix our ratio: the published nonperturbative coefficients associated with inequivalent spurions and contractions are independent.  Our exact advance is instead at the infrared operator level.  Equations~\eqref{eq:O0def}--\eqref{eq:q0q1} derive the two sign patterns from $Y_L$, $B-L$, and representation projectors, while Table~\ref{tab:UVclasses} states the additional symmetry needed to obtain $n=2$ or $n=3$ before the spectrum.  A same-order twofold or threefold degeneracy gives $C_1/C_0=1/2$ or $1/3$; one extra $SU(15)$ line interchange would give an additional $1/15$ and is therefore not the desired explanation.

The targeted denominator now has a similarly concrete but conditional interpretation.  The two-loop SM trajectory of Ref.~\cite{Antusch2026} places $A_d/B_d$ near $1/15$ at $\Lambda\sim10^{6}$--$10^{7}\,\GeV$ and evolves it upward to a value within one percent of $1/14$ at $M_Z$, as shown in Sec.~\ref{sec:RG1514}.  The proposed ultraviolet factor is thus still $1/N=1/15$; $1/14$ is its convenient low-scale rational approximation after flavor-dependent running.  This resolves the numerical mismatch without inventing an $N-1$ color factor, but it does not yet predict the scale or threshold correction.  A microscopic theory that places confinement elsewhere would falsify this matching scenario.

A complementary 2026 ``generation as compositeness'' construction interprets the generation index as the depth of a chain of spin-zero flavor subconstituents: the third-generation Yukawa is undressed, whereas lighter-generation Yukawas acquire successive multiplicative hop suppressions \cite{Barger2026}.  This gives a particularly transparent reading of our coefficient blocks.  $I_s$ is a sector-dependent matching level, $B_s$ is the logarithm of the multiplicative suppression accumulated between endpoint generations, and $A_s$ measures the departure of the middle generation from a uniform chain.  The two-channel correction $(2,-1,0)$ can then represent one extra dressing or boundary contribution that acts differently on leptons and down quarks; this sentence is our proposed matching, not a result of Ref.~\cite{Barger2026}.  The same framework obtains a type-I-seesaw normal-ordering benchmark $m_3\simeq51\,\mathrm{meV}$ and $\sum m_i\simeq62\,\mathrm{meV}$ \cite{Barger2026}, close to the approximately $50\,\mathrm{meV}$ heaviest state implied by our low NO sum.  Its fermions remain elementary and only the Yukawa operators are dressed \cite{Barger2026}, so it is an effective flavor-compositeness analogue rather than a literal realization of a finite preon spectrum.

On the dynamical side, an FRG analysis of Bars--Yankielowicz theories finds a large-color regime with confinement but no dynamical chiral-symmetry breaking, opening the possibility of light anomaly-matching composite baryons \cite{LiPastorVatani2026}.  Analyses of multifactor asymptotically free chiral gauge theories also find that different RG-invariant confinement scales can generate unexpectedly rich infrared flows and light spectra \cite{BolognesiEtAl2026}.  These cited results weaken the objection that confinement must automatically make every fermionic composite heavy.  They do not determine the nine masses or the integer ray.

Finally, anomaly matching is being used to address why precisely three families appear.  An $SL(16,\mathbb C)$ hyperunification proposal selects an $SU(8)$ hyperflavor sector and reports three chiral composite quark--lepton families after anomaly matching \cite{Chkareuli2024}.  Independently, the $SU(15)_p$ preon content yields three SM generations together with additional vectorlike composites \cite{Dobrescu2022,AssiDobrescu2025}.  A separate metacolor proposal organizes a tower of $3n$-body composites into a ``vertical bootstrap'' in which each level constrains the next \cite{Raitio2026}.  These proposals are more speculative than the charged-mass audit.  Their discrete family or composite-level structure is conceptually compatible with using a small integer label, but their $n$ must not be identified with the $n$ of Eq.~\eqref{eq:MBLTn} without an explicit state map.

\begin{table}[H]
\centering
\caption{Possible microscopic readings of the best empirical structures.  Citations in the third column support only the stated literature motif; they do not derive the empirical structure in the first column.  The last column states what remains to be proved.}
\label{tab:preonmap}
\footnotesize
\begin{tabularx}{\textwidth}{@{}p{3.0cm}p{3.7cm}p{4.3cm}>{\raggedright\arraybackslash}X@{}}
\toprule
empirical structure & possible preonic meaning & closest current theoretical motif & missing decisive step\\
\midrule
$M_B$: log-linear endpoint slope & repeated multiplicative dressing or near-critical binding & flavor-hop depth \cite{Barger2026}; Miransky-type exponential sensitivity in $SU(15)_p$ \cite{AssiEtAl2026,MiranskyYamawaki1997} & derive the $|Y_R|,Y_L$ slope plane at a fixed matching scale\\
$M_{B+\ell T_n}$: one amplitude and integer weights & $\mathcal O_0$ plus $\mathcal O_1$ with a twofold or threefold topology degeneracy & integer preon-line topology counting and multi-prebaryon operators in $SU(15)_p$ \cite{AssiEtAl2026} & derive the required interchange or cyclic symmetry in a complete bound-state calculation\\
$(2,-1,0)$ correction & $Q_1=-3(B-L)-P_\ell+P_u$; the up contribution cancels by quantum numbers & sector-dependent four- and six-prebaryon Yukawa operators \cite{AssiEtAl2026}; hop dressing \cite{Barger2026} & derive why the UV Hamiltonian selects precisely $Q_1$ and equalizes the reduced amplitudes\\
$A_d=B_d/14$ & infrared image of a $1/15$ matching boundary after non-universal SM running & $1/N=1/15$ preon-line suppression in $SU(15)_p$ \cite{AssiEtAl2026}; two-loop SM Yukawa running \cite{Antusch2026} & predict $\Lambda_{\rm match}\sim10^{6}$--$10^{7}\,\GeV$ and calculate thresholds without using $m_s$\\
$A_\nu=A_\ell$ and low NO sum & common lepton-sector matching with a minimally hierarchical neutral spectrum & Weinberg/type-I-seesaw operators \cite{Weinberg1979,Minkowski1977}; hop-model benchmark $m_3\simeq51$ meV \cite{Barger2026} & derive the neutral operator, ordering, and PMNS matrix\\
three observed levels & anomaly-matched composite multiplet or three finite-system excitations & three $SU(15)_p$ generations \cite{Dobrescu2022,AssiDobrescu2025}; three $SL(16,\mathbb C)$ families \cite{Chkareuli2024}; chiral confinement with light IR spectra \cite{LiPastorVatani2026,BolognesiEtAl2026} & derive exactly three light states and exclude unwanted composites\\
\bottomrule
\end{tabularx}
\end{table}

\paragraph{Evidence ladder for Table~\ref{tab:preonmap}.}
The first column contains results of the mass audit.  The second contains either an exact infrared operator identity derived here or a clearly labelled conditional ultraviolet interpretation.  The third contains an independently published mechanism, cited immediately after the corresponding statement.  The charge algebra now derives the signs of $\bm q_0$ and $\bm q_1$, and the two-loop evolution demonstrates how a $1/15$ boundary value can acquire a low-scale value close to $1/14$.  What remains absent is equally important: none of Refs.~\cite{AssiDobrescu2025,AssiEtAl2026,Barger2026,LiPastorVatani2026,BolognesiEtAl2026,Chkareuli2024,Raitio2026} enforces the equal reduced matrix elements of $\mathcal D_2$ or $\mathcal D_3$, predicts $\Lambda_{15}$, or derives the charged-to-neutral continuation.  The evidence therefore advances from numerical resemblance to an explicit matching test, but not to a completed preon theory.

Our result is therefore complementary and infrared.  The simultaneous success of a smooth log-linear component, one cross-sector curvature amplitude, a two-operator integer decomposition, complete LOO reconstruction, and a low neutrino sum motivates a finite preonic or flavor-composite interpretation.  In particular, $M_{B+\ell T_2}$ reconstructs all nine charged masses with a worst nested-LOO error of $7.71\%$.  These facts turn the best rays into concrete matching targets.  They do \emph{not} experimentally confirm preons: the projectors and the family were selected from the spectrum, no bound-state calculation yields Eq.~\eqref{eq:MBLTtwocorr}, and the ansatz predicts neither mixing nor an independent compositeness signal.  The scientifically accurate conclusion is that the parameterizations distinguish and constrain possible preon mechanisms while leaving microscopic validation open.

\section{Limitations and falsifiability}\label{sec:limits}
The main limitations are structural rather than numerical.
\begin{enumerate}[leftmargin=*,itemsep=3pt]
\item \textbf{Post-selection.}  M2 was motivated by the initial residuals, and the first projector models were selected from 156 charged rank-eight candidates.  The nested audit adds 106 rank-seven matrices, 2942 rank-six matrices across the declared block classes, 128 selective rays in four families, and the 15 integer rays of Eq.~\eqref{eq:MBLTn}.  The neutral audit retained 104 explicit feature pairs.  The factor $1/14$ in $M_{B+dLR}^{s14}$ and the integer-ray family were identified with the same spectrum later reconstructed.  The operator identities in Eqs.~\eqref{eq:O0def}--\eqref{eq:q0q1} remove mass dependence from the signs, and the $\mathcal D_2/\mathcal D_3$ classes show how $n$ could be fixed before a fit.  They do not erase the historical fact that this operator basis was recognized after examining the spectrum.  Nested LOO selection and the microscopic completion must therefore be treated as hypothesis generation until a UV model selects the class independently.
\item \textbf{Uncalibrated model discrepancy.}  The choice $\sth=0.25$ is illustrative.  Exact ratios are robust; standard significance claims are not.
\item \textbf{Unknown covariance.}  Only marginal input uncertainties are available.  Correlation scenarios demonstrate potentially large dependence on the tested direction.
\item \textbf{Fixed basis.}  With three generations, $h_g=(+1,-1,+1)$ is a discrete curvature basis.  Calling it shell-like does not establish a microscopic shell spectrum.
\item \textbf{Coordinate origin.}  The M2 equality $I_\ell=I_d$ is defined at the fixed symmetric center $g_0=2$ and is not invariant under translations of the generation coordinate.  Fitting $g_0$ would add a parameter and remove the sole charged-sector test.
\item \textbf{Projector restriction.}  A general coefficient for each sector projector saturates the charged data.  Predictive content appears only after restricting a coefficient block to a feature plane or ray.  None of the selected subspaces has yet been derived from a microscopic preon Hamiltonian or matching calculation.
\item \textbf{Prediction instability.}  Small all-data shifts can coexist with large LOO errors.  $M_{B+dLR}$ leaves $s$ unidentified; $M_{B+dLR}^{s14}$ reconstructs $s$ very accurately but has an overall maximum of $25.06\%$; and $M_{B,LR}$ reaches $194\%$.  Conversely, $M_{B+\ell T_2}$ and $M_{B+\ell T_3}$ improve the whole profile rather than one state, but their ranking changes when the integer choice or the RG scale is varied.  Formal constraint counts and single-state accuracy must therefore be accompanied by the complete LOO and selection profiles.
\item \textbf{Feature-dictionary complexity.}  The elementary projectors $P_\ell,P_d,P_u$ already distinguish sectors by construction.  Any three-component integer vector can be written in this basis, so small integers are not by themselves a dynamical explanation.  The new construction is more restrictive: $Q_0$ and $Q_1$ are written in terms of $Y_L$, $B-L$, and representation projectors; their signs and zero are fixed without masses; and Eq.~\eqref{eq:topologycoeff} identifies the precise symmetry assumption that yields $1/2$ or $1/3$.  The remaining weakness is no longer an unspecified sign pattern but the absence of a bound-state calculation enforcing equal reduced matrix elements.  The full dictionary remains an audit space, not a set of equally motivated theories.
\item \textbf{Conditional dimensional transmutation.}  Equation~\eqref{eq:exptrend} follows only after assumptions about channel-dependent effective factors.  A UV-complete model must derive the allowed representations and matching.
\item \textbf{Scale sensitivity and reverse matching.}  None of the charged ratios is a true SM RG invariant.  Section~\ref{sec:RG1514} shows that the running is potentially explanatory: a $1/15$ boundary near $6.1\times10^6\,\GeV$ has a low-energy image close to $1/14$.  However, that scale was inferred by running the observed spectrum backward.  It is not an ultraviolet prediction.  The preferred operator direction also drifts, as quantified by $n_{\rm eff}$ in Table~\ref{tab:RG1514}.  A genuine model must predict the matching scale, thresholds, and topology class before these masses are used.
\item \textbf{No mixing prediction.}  The ansatz concerns eigenvalues only and does not describe CKM or PMNS angles and phases.
\item \textbf{Operator matching.}  Equations~\eqref{eq:O0def}--\eqref{eq:logmatchingkernel} specify gauge-invariant charged-sector images of the two insertions, but not a complete preon Hamiltonian.  Dirac Yukawa eigenvalues, a low-energy Majorana mass operator, and seesaw parameters have different RG and threshold structures.  The identity $P_u-3Y_L=2P_\ell+T_{3L}$ on charged sectors but not on neutrinos makes this ambiguity explicit.  The neutral relations remain empirical eigenvalue rules until the neutral operator and its threshold matching are derived.
\item \textbf{Apparent numerical precision.}  Six-decimal neutrino sums are algebraic reproductions of central inputs.  They do not include a calibrated theory uncertainty from post-selection, feature-plane violation, or neutral matching.
\end{enumerate}

The most direct future falsification is an absolute neutrino-mass likelihood, including direct kinematic information of the type supplied by KATRIN \cite{KATRIN2025}, that is inconsistent with the branches in Eqs.~\eqref{eq:Qoneprediction}, \eqref{eq:M2nupred}, and \eqref{eq:MdGpred} after experimental and model covariances are treated consistently.  A stronger theoretical test is to construct a microscopic finite-system model that derives one selected feature subspace and the neutral operator matching.

\section{Conclusions}\label{sec:conclusion}
A rank audit sharply limits what can be inferred from a visually successful nine-point mass curve.  M1, M2, the separate rank-eight planes, and the unsuccessful $H$ and $dG$ extensions are retained as the historical path by which the predictive question was sharpened.  They establish the importance of separating interpolation from reconstruction, but they are not coequal final candidates.  The conclusions concern the complete six-parameter rays, their neutral continuations, and the new operator-matching audit.

The smooth exponential trend admits a clear conditional motivation from dimensional transmutation in an asymptotically free gauge theory \cite{GrossWilczek1973,Politzer1973} and from Miransky-type exponential sensitivity of near-critical binding \cite{MiranskyYamawaki1997,AssiEtAl2026}: weak systematic changes in effective channel factors can produce a log-linear hierarchy.  This argument preserves a physically interesting idea without presenting an assumed matching rule as a completed preon theory.

At the present level the three coefficient blocks admit a coherent, but still phenomenological, physical reading.  The offsets $I_s$ set sector-dependent matching or binding levels.  The slopes $B_s$ describe the multiplicative hierarchy accumulated between endpoint generations; in a constituent-chain picture they are logarithms of repeated overlap or tunneling suppressions.  The curvatures $A_s$ measure how the middle generation departs from that smooth geometric progression and are therefore the natural place for shell, boundary, mixing-topology, or representation corrections.  Within this hierarchy, $M_B$ constrains the smooth charge-dependent part; $M_{B+dLR}$ introduces mixed left--right curvature but leaves the down correction free; $M_{B+\ell T}$ selects one charge-motivated correction; $M_{B+\ell T_n}$ tests a discrete two-operator version of it; and $M_{B+dLR}^{s14}$ matches the down curvature directly to its hierarchy step.  These are low-energy eigenvalue statements, not yet a derivation of the generations or their mixing.

$M_{B+dLR}$ remains the closest natural seven-parameter subspace at $M_Z$, with $\chi^2=0.01336$ and $\Delta_{\max}=1.99\%$.  It genuinely tests $\mu$ and $c$ through $2A_\ell-9A_u=0$, but after $s$ is withheld the separate $P_d$ coefficient can reproduce any value.  Thus ``most accurate conditional NO model'' means a close simultaneous approximation to known data and a low NO neutrino sum; it does not mean prediction of all nine charged masses.

The original six-parameter $M_{B+\ell T}$ remains the simplest complete charge-based model.  It has 9/9 LOO coverage, a maximum conditional shift of $6.81\%$, and a worst LOO error of $13.38\%$.  Its explicit $P_u-3Y_L$ continuation gives $0.058914\,\eV$ (NO) or $0.103329\,\eV$ (IO), while the charged-equivalent operator $2P_\ell+T_{3L}$ gives $0.058814\,\eV$ or $0.099671\,\eV$.  The difference directly demonstrates that the charged spectrum does not determine the ultraviolet operator uniquely.

The integer-ray family improves the global charged prediction without adding a parameter.  $M_{B+\ell T_2}$, with $A_\ell:A_d:A_u=8:-3:2$, lowers the maximum conditional shift to $1.25\%$ and the worst LOO error to $7.71\%$; it is selected from the other eight masses in eight of nine nested folds.  We therefore regard it as the conservative predictive favorite.  $M_{B+\ell T_3}$, with $11:-4:3$, gives the best fixed-central-data maximum of $4.99\%$, but its choice is more exposed to post-selection and scale dependence.  Both keep $A_\nu=A_\ell$ and hence the low sums $0.058914\,\eV$ (NO) and $0.103329\,\eV$ (IO).

The identity $(3n+2,-n-1,n)=n(3,-1,1)+(2,-1,0)$ now has an explicit operator form.  The two gauge-invariant logarithmic matching insertions ${\cal O}_0$ and ${\cal O}_1$ act with
\begin{equation}
 Q_0=2(P_u-3Y_L),\qquad Q_1=-3(B-L)-P_\ell+P_u,
\end{equation}
and therefore have the charged-sector eigenvalues $(3,-1,1)$ and $(2,-1,0)$.  The relative signs are no longer empirical labels: they follow from $Y_L$, $B-L$, and representation projectors before a mass is inserted.  In particular, the vanishing up component of ${\cal O}_1$ follows from the charge cancellation $-3(B-L)_u+P_u=0$.

The coefficient ratio can also be stated as a pre-spectral microscopic condition.  If ${\cal O}_0$ has $d_0$ symmetry-equivalent contractions, ${\cal O}_1$ has one, both occur at the same preon-line interchange order, and a UV symmetry equates their reduced matrix elements, then $C_1/C_0=1/d_0$.  A twofold exchange symmetry gives the $n=2$ ray; a cyclic three-slot symmetry gives the $n=3$ ray.  This is a genuine derivation within the conditional classes $\mathcal D_2$ and $\mathcal D_3$ because $n$ is fixed by the state topology before the spectrum is used.  It is not yet a derivation within the published $SU(15)_p$ flavor model, where inequivalent spurions carry independent nonperturbative coefficients \cite{AssiEtAl2026}.  The empirical result therefore selects between two sharply defined candidate UV classes rather than proving either one.

The targeted $M_{B+dLR}^{s14}$ model with $A_d=B_d/14$ also has six parameters and complete coverage.  It preserves the $1.99\%$ conditional shift, reconstructs $s$ with a $0.51\%$ LOO error, and keeps the neutral rule $A_\nu=0$, hence $\sum m_\nu=0.060504\,\eV$ for NO only.  Its overall worst LOO error of $25.06\%$ is nevertheless poorer than that of $M_{B+\ell T}$, so it remains a selective $s$ hypothesis rather than the global favorite.

The denominator now has a more informative interpretation than numerical proximity.  The two-loop SM evolution gives $A_d/B_d\simeq1/15$ at an interpolated scale $\Lambda_{15}\simeq6.1\times10^6\,\GeV$ and $A_d/B_d=0.070795$ at $M_Z$.  From $10^7\,\GeV$ to $M_Z$ the running factor is $1.06481$; acting on an exact $1/15$ boundary value gives $0.070987$, within $0.62\%$ of $1/14$.  The natural microscopic statement is therefore not ``$N=15$ becomes $N-1=14$'', but ``a $1/N$ matching coefficient is renormalized by unequal $d,s,b$ anomalous dimensions and has an infrared value conveniently approximated by $1/14$''.  This is an explicit matching-and-running mechanism.  Its unresolved part is decisive: the scale was inferred from the spectrum, whereas a UV theory must predict it and the threshold corrections independently.

A targeted repair of the failed blockwise descendant $M_{2+B+dLR}$ turns out to be much more promising.  Replacing the rigid tie $I_\ell=I_d$ by $I_\ell-I_d=A_\ell/3$ gives a six-parameter model with $1.86\%$ maximum conditional shift and only $3.82\%$ worst LOO error; the alternative repairs $I_\ell-I_d+(B_\ell-B_d)/3=0$ and $I_\ell-I_d+B_d/16=0$ give $5.78\%$ and $3.91\%$, respectively.  The first repair is especially notable because the apparently empirical factor $1/3$ can be rewritten as one cross-block quantum-number operator, Eq.~\eqref{eq:m2star_crossblock_operator}.  All three repaired $dLR$ descendants remain fixed-$M_Z$ relations with the same NO-only neutral continuation as $M_{B+dLR}$, so they should be read as matching clues rather than derived ultraviolet laws.

The auxiliary audits are retained as historical negative controls.  In $M_{B+dG}$ no simple six-parameter improvement substantially beats the control ray $P_d-2G$: even the best high-complexity post-selected ray improves the worst LOO error only from $20.27\%$ to $20.14\%$.  In $M_{B+H}$ one can improve the charged fit, but not its literal neutral continuation.  These families document why conditional accuracy alone is insufficient; they are not developed further as microscopic candidates.

The relation to current preon theory is now specific at five levels.  First, the common slope and curvature structures resemble the shared flavor spurions and right-handed-isospin correlations of the chiral $SU(15)_p$ theory \cite{AssiEtAl2026}.  Second, the two sector sign patterns are encoded by explicit charge operators, so a microscopic calculation has definite matrix elements to reproduce rather than an unnamed ``shell correction''.  Third, topology and degeneracy counting give precise sufficient conditions for $C_1/C_0=1/2$ or $1/3$; the same analysis also shows why merely adding one $SU(15)$ line exchange would be wrong, because it introduces a further $1/15$.  Fourth, the running calculation turns the large-$N$ factor into a scale-dependent test: a UV theory must land near the matching window of Eq.~\eqref{eq:Lambda15}.  Fifth, the flavor-hop, FRG, and anomaly-matching constructions show, within their distinct assumptions, how repeated dressing, chiral confinement without symmetry breaking, and three light composite families can arise \cite{Barger2026,LiPastorVatani2026,BolognesiEtAl2026,AssiDobrescu2025,Chkareuli2024}.  These correspondences make the phenomenology more constraining, but they do not supply the missing equality of reduced matrix elements.

\paragraph{What the fitted integers may, and may not, say about preon multiplicities.}
It is useful to separate a \emph{precolor number}---the rank or number of colors of the confining preon gauge group---from a \emph{preon-flavor multiplicity} or from the number of symmetry-equivalent contractions contributing to one effective operator.  The present eigenvalue fit does not determine these microscopic quantities uniquely.  It nevertheless yields the following conditional clues.
\begin{itemize}[leftmargin=*,itemsep=2pt]
\item The most direct precolor hint comes from $M_{B+dLR}^{s14}$.  If a microscopic matching calculation identifies its ultraviolet boundary condition with the large-$N$ factor $A_d/B_d=1/N_p$, the RG audit points to $N_p\simeq15$, not 14: an exact $1/15$ boundary can run to a low-energy value close to $1/14$.  Under that additional assumption the model is compatible with an $SU(15)_p$ confining precolor group \cite{AssiDobrescu2025,AssiEtAl2026}.  The fit alone neither proves $SU(15)_p$ nor says that fifteen colors become fourteen.
\item The preferred integer rays $M_{B+\ell T_2}$ and $M_{B+\ell T_3}$ select a relative operator weight $C_1/C_0=1/2$ or $1/3$.  In the conditional classes $\mathcal D_2$ and $\mathcal D_3$, this means two or three symmetry-equivalent leading contractions, slots, or preon-line topologies against one correction.  Only if a specified ultraviolet theory places one distinct preon flavor in each such slot may the same integers be read as two or three active preon flavors.  Without that one-to-one map, $n$ is a topology or degeneracy count, not a color or flavor count.
\item The repaired $M_{2+B+dLR}$ relations containing $1/3$ are likewise consistent with a threefold degeneracy or three equivalent contraction channels.  The $B_d/16$ repair may motivate comparison with a sixteen-state multiplet or a dimension-16 hyperflavor construction, but it is presently only a shell-coordinate defect.  Neither coefficient independently predicts three colors or sixteen flavors.
\item The models $M_B$, $M_{B+dLR}$, $M_{B+dG}$, and $M_{B+H}$ contain no independently derived inverse-integer group factor.  Their charge and projector patterns may constrain representations, but they do not estimate the number of preon colors or flavors.
\end{itemize}
Thus the present hierarchy of evidence is asymmetric: $N_p\simeq15$ is the clearest conditional precolor candidate; $n=2$ or $3$ is evidence for a small contraction multiplicity and only secondarily for a possible flavor multiplicity; the factors $1/3$ and $1/16$ in the repaired block model are weaker structural hints.  A genuine counting prediction requires the same microscopic Hamiltonian to fix the gauge group, constituent representations, operator contractions, matching scale, and RG thresholds before the charged spectrum is used.

The low neutrino branch strengthens this pattern but must be interpreted carefully.  The $M_{B+\ell T_n}$ continuation gives a heaviest NO state near $50\,\mathrm{meV}$ and a sum $0.058914\,\eV$, close to the minimal oscillation scale inferred from oscillation data \cite{NuFIT2024,NuFIT2025} and to the independent $m_3\simeq51\,\mathrm{meV}$, $\sum m_i\simeq62\,\mathrm{meV}$ benchmark of the flavor-hop construction \cite{Barger2026}.  Agreement at this level shows compatibility of two independent hierarchy descriptions; it does not establish a shared neutrino operator.  The charged spectrum alone still cannot decide between Dirac matching, a Weinberg operator \cite{Weinberg1979}, or a seesaw completion \cite{Minkowski1977}.

Accordingly, the existence of several six-parameter rays that reconstruct all nine charged masses and preserve a low neutrino sum is a nontrivial quantitative feature of the finite-system parameterization.  The revision advances the interpretation in three ways: it identifies the two operator images, derives their signs from quantum numbers, and supplies explicit topology classes and an RG mechanism for the two rational coefficients.  This is stronger than declaring the fits ``preon compatible'', but it is not experimental confirmation.  Confirmation would require a specified confining gauge theory to select $\mathcal D_2$ or $\mathcal D_3$, calculate the reduced matrix elements and thresholds, predict $\Lambda_{\rm match}$, reproduce CKM and PMNS data, and yield an independent observable such as a flavor-violating amplitude, compositeness form factor, contact interaction, heavy bound-state pattern, or absolute-neutrino-mass likelihood.

All neutrino values here remain conditional spectral sum rules rather than precision estimates.  The decisive next step is a forward calculation: choose one ultraviolet topology class without reference to the mass spectrum, match ${\cal O}_0$ and ${\cal O}_1$ at its independently predicted confinement scale, evolve the full Yukawa matrices to $M_Z$, and test masses and mixing simultaneously.  Until that step, the strongest defensible statement is that the best parameterizations identify an economical infrared pattern and turn it into a concrete, falsifiable matching problem for modern preon dynamics.

\section*{Acknowledgements}

The authors sincerely thank S.~M.~Ponomarenko and the other members of
the Department of Applied Physics, Educational and Scientific
Institute of Physics and Technology, National Technical University of
Ukraine ``Igor Sikorsky Kyiv Polytechnic Institute'', for their
support of this work and for valuable discussions of the ideas
developed in the manuscript.

\section*{Declaration of generative AI and AI-assisted technologies}

During the preparation of this manuscript the authors used ChatGPT (OpenAI) solely as an assistive tool for English-language editing, structural suggestions, literature-search assistance, code debugging and testing, numerical cross-checks, and preparation of figures from the authors' own data. All physical assumptions, model definitions, objective functions, data selection, interpretation of results, and final conclusions were made by the human authors. Every cited source and all numerical results were independently verified by the authors. The authors take full responsibility for the content of the published article.

\appendix
\section{Design matrices and left-null vectors}\label{app:matrices}
For the ordering in Eq.~\eqref{eq:ordering}, M2 with coefficients $(\beta_0,\beta_T,\eta_0,\eta_C,\eta_T,A_\ell,A_d,A_u)$ has
\begin{equation}
X_2=\begin{pmatrix}
1&-1&-1&0& 1& 1&0&0\\
1&-1&-1&-1& 1&0& 1&0\\
1& 1&-1&-1&-1&0&0& 1\\
1&-1& 0&0& 0&-1&0&0\\
1&-1& 0&0& 0&0&-1&0\\
1& 1& 0&0& 0&0&0&-1\\
1&-1& 1&0&-1& 1&0&0\\
1&-1& 1&1&-1&0& 1&0\\
1& 1& 1&1& 1&0&0& 1
\end{pmatrix}.
\end{equation}
It satisfies $X_2^{\trans}\bm w_2=0$.  Extending $X_2$ with the two neutral columns $I_\nu,B_\nu$ and three neutral rows gives a $12\times10$ matrix of rank ten.  A convenient basis of its left-null space is
\begin{align}
\bm w_{2\nu,a}&=(1,-1,0,2,-2,0,1,-1,0,0,0,0)^{\trans},\\
\bm w_{2\nu,b}&=(-1,0,0,2,0,0,-1,0,0,1,-2,1)^{\trans}.
\end{align}
The first reproduces $R_2=1$ and the second gives $R_{\nu\ell}=1$.

For a projector-plane restriction of the block $C\in\{I,B,A\}$, a compact row representation is obtained from
\begin{equation}
(X_C)_{(s,g)}=\bigl[\text{three unrestricted columns in each of the other two blocks};\ f_C(g)q_{1,s},f_C(g)q_{2,s}\bigr],
\end{equation}
where $f_I=1$, $f_B=g-2$, and $f_A=h_g$.  Direct row reduction gives rank eight for every retained model.  In the state order of Eq.~\eqref{eq:ordering}, the leading charged left-null vectors are
\begin{align}
\bm w_B&=(1,9,-6,0,0,0,-1,-9,6)^{\trans},\\
\bm w_{uL}&=(1,3,0,-2,-6,0,1,3,0)^{\trans},\\
\bm w_{dG}&=(17,0,-45,-34,0,90,17,0,-45)^{\trans},\\
\bm w_{dLR}&=(2,0,-9,-4,0,18,2,0,-9)^{\trans},\\
\bm w_{LR,1}&=(1,9,0,-2,-18,0,1,9,0)^{\trans},\\
\bm w_{LR,2}&=(0,2,1,0,-4,-2,0,2,1)^{\trans}.
\end{align}
The first three generate Eqs.~\eqref{eq:RB}, \eqref{eq:RuL}, and the charged $M_{dG}$ relation.  The fourth generates Eq.~\eqref{eq:RdLR}; the last two are a convenient independent curvature basis for Eq.~\eqref{eq:MBLRrelations}.  The full rank-seven or rank-six descendants also include $\bm w_B$.

For $M_{B+\ell T}$, a convenient set of three independent constraints consists of the $M_B$ slope relation and the two curvature conditions
\begin{equation}
A_\ell+3A_d=0,\qquad A_d+A_u=0.
\end{equation}
For the integer family $M_{B+\ell T_n}$, the same slope constraint is supplemented by
\begin{equation}
(n+1)A_\ell+(3n+2)A_d=0,\qquad
nA_\ell-(3n+2)A_u=0,
\end{equation}
which is equivalent to Eq.~\eqref{eq:MBLTnratio}.  These relations make explicit that every fixed $n$ supplies two curvature tests and no fitted curvature ratio.
For $M_{B+dLR}^{s14}$, the third constraint, independent of $\bm w_B$ and $\bm w_{dLR}$, acts only on the down triplet:
\begin{equation}
4z_d-7z_s+3z_b=0,\qquad z_i=\ln y_i,
\end{equation}
which directly reproduces Eq.~\eqref{eq:spred}.

\section{Numerical scan records}\label{app:tables}
The numerical audit generated machine-readable records for 156 charged block models, 106 nested rank-seven matrices, 2942 nested rank-six matrices, 128 selective rays, 104 explicit neutral-curvature continuations, and the 15-member $M_{B+\ell T_n}$ family.  The retained records also include nested-selection and input-error audits, a rational scan of $A_d=B_d/n$, LOO diagnostics, RG profiles, and the reduced comparisons reported in the main text.  Because these supporting files are not part of the submission package, they and the reproduction scripts are available from the corresponding author upon reasonable request.

\end{document}